\documentclass[traditabstract]{aa} 

\usepackage{caption}
\usepackage{subcaption}
\usepackage{epstopdf}
\usepackage{amsmath}
\usepackage[english]{babel}
\usepackage{amssymb}
\usepackage{wrapfig}
\usepackage{pdflscape}
\usepackage{lscape}
\usepackage{longtable}
\usepackage{appendix}
\usepackage{xcolor}
\usepackage{textcomp}
\usepackage{multirow}
\bibpunct{(}{)}{;}{a}{}{,}
\usepackage{array}
\usepackage{ragged2e}
\usepackage{adjustbox}

\usepackage{makecell}
\usepackage{graphicx}
\usepackage{natbib}
\usepackage{txfonts}
\usepackage{hyperref}
\begin{document}

   \title{The host galaxy of low-luminosity compact sources}

   \author{A. Vietri
        \inst{1,2}\thanks{amelia.vietri@phd.unipd.it}
          \and
          M. Berton\inst{2}\thanks{marco.berton@eso.org}
          \and
          E. J\"arvel\"a \inst{3,4}\thanks{Dodge Family Prize Fellow in The University of Oklahoma}
          \and
          M. Kunert-Bajraszewska \inst{5}
          \and
          S. Ciroi \inst{1}
          \and
          I. Varglund \inst{6,7}
          \and 
          B. Dalla Barba \inst{2,8,9}
          \and
          E. Sani \inst{2} 
          \and
          L. Crepaldi \inst{1,2,10,11}
          }

   \institute{
   $1$ Dipartimento di Fisica e Astronomia ``G. Galilei", Universit\`a di Padova, Vicolo dell'Osservatorio 3, 35122, Padova, Italy; \\
   $^2$ European Southern Observatory, Alonso de C\'ordova 3107, Casilla 19, Santiago 19001, Chile;\\
   $^3$ Homer L. Dodge Department of Physics and Astronomy, The University of Oklahoma, 440 W. Brooks St., Norman, OK 73019, USA; \\
   $^4$ European Space Agency (ESA), European Space Astronomy Centre (ESAC), Camino Bajo del Castillo s/n, 28692 Villanueva de la Ca\~nada, Madrid, Spain; \\
   $^5$ Institute of Astronomy, Faculty of Physics, Astronomy and Informatics, NCU, Grudziądzka 5/7, 87-100, Toruń, Poland \\
   $^6$ Aalto University Metsähovi Radio Observatory, Metsähovintie 114, FI-02540 Kylmälä, Finland \\
   $^7$ Aalto University Department of Electronics and Nanoengineering, P.O. Box 15500, FI-00076 AALTO, Finland \\
   $^8$ Dipartimento di Scienza e Alta Tecnologia, Università dell'Insubria, Via Valleggio 11, 22100 Como, Italy\\
   $^9$ Osservatorio Astronomico di Brera, Istituto Nazionale di Astrofisica (INAF), 23807 Merate, Italy\\
   $^{10}$ INAF - Osservatorio Astronomico di Cagliari, Via della Scienza 5, 09047 Selargius (CA), Italy\\
   $^{11}$ Departamento de Física y Astronomía, Universidad de La Serena, Av. Cisternas 1200 N, La Serena, Chile
             }

   \date{Received ...; accepted ...}

 
  \abstract
   {The term `active galactic nuclei' (AGN) subtends a huge variety of objects, classified on their properties at different wavelengths. Peaked sources (PS) represent a class of AGN at the first stage of evolution, characterised by a peaked radio spectrum. Among these radio sources, low-luminosity compact (LLC) sources can be identified as PS accreting with a high Eddington rate, harbouring low-power jets and hosting low-mass black holes. These properties are also shared by narrow-line Seyfert 1 galaxies (NLS1s). In 2016, LLCs were hypothesised to be the parent population of NLS1s with a flat radio spectrum (F-NLS1s), suggesting the former to be the same objects as the latter, seen at higher inclination. Based on radio luminosity functions and optical spectra analysis, 10 LLCs were identified as valid candidates for F-NLS1s. To account for the missing puzzle piece, verifying if these LLCs could be hosted in late-type galaxies as NLS1s, we performed the photometric decomposition of their Pan-STARRS1 images in all five filters. We used the 2D fitting algorithm GALFIT for the single-band analysis, and its extension GALFITM for the multi-band analysis. Considering that the morphological type and the structural parameters of the host can be dependent on the wavelength, we found six out of ten LLCs hosted in late-type galaxies, probably with pseudo-bulges, three point-like sources and one object of uncertain classification. Although this study is based on a small sample, it represents the first morphological analysis of LLC host galaxies. The results confirm the trend observed in NLS1s, indicating late-type/disc-like host galaxies for LLCs and supporting the validity of the parent population scenario.}

   \maketitle
%
\newcommand{\kms}{km s$^{-1}$}
\newcommand{\ergs}{erg s$^{-1}$}
\newcommand{\chired}{$\chi^2_\nu$}
\section{Introduction}
Among jetted active galactic nuclei (AGN), peaked sources (PS) are compact and powerful radio sources, often considered an early stage of radio galaxy (RG) evolution \citep{Odea98}, that is, sources which will eventually grow into the giant RGs often labelled as Fanaroff-Riley (FR) I/II \citep{Fanaroff74}. PS are typically characterised by a convex radio spectrum, whose peak position anticorrelates with their linear size \citep{Odea97}, which is, by definition, smaller than 15 kpc \citep{Odea98}. This means that their relativistic jets are usually confined within their host galaxy, and can often interact with the interstellar medium \citep{Orienti16}. The peaked spectrum is a byproduct of the small jet size, either due to
synchrotron self-absorption or free-free absorption. However, above the peak, the jet emission is optically thin, and the spectral slope is that of typical synchrotron radiation. \par

The label PS, however, is a large umbrella that covers a variety of different objects. Some of them may not be genuinely young, since PS can include also frustrated sources or AGN with intermittent jet activity \citep{Odea21}. Furthermore, different classifications exist for PS, depending on the position of the peak in the radio spectrum. Compact steep-spectrum sources (CSS) peak in the MHz range, gigahertz-peaked sources (GPS) in the GHz range, while high-frequency peakers above $\sim$5 GHz. Among all the subclass of PS, low-luminosity compact (LLC) sources, characterised by a luminosity \textit{L}$<10^{26}$ W Hz$^{-1}$ at 1.4 GHz \citep{Kunert10a}, are one of the most interesting. As for other RGs, LLCs can be divided into high-excitation (HE) and low-excitation (LE) sources, based on the ionisation degree measured in the optical spectrum \citep{Kunert10b}. Many LLC may be short-lived objects that will evolve into RGs only after several activity episodes \citep{Kunert10a}, of about 10$^3$-$10^4$ yr, separated by periods of 10$^4$-$10^6$ yr when the jet is switched off \citep{Czerny09}. On the other hand, recent findings also prove that some compact sources keep their small size during their whole life cycle and do not evolve into larger jetted objects \citep{Readhead24, Kiehlmann24}. \par

An explanation for the low luminosity of LLCs could be their black hole mass. Indeed, it is known that the jet power scales non-linearly with the mass of the central engine \citep{Heinz03}, and according to recent studies, also with the magnetic flux, through a relation that still needs to be explored \citep{Chamani21}. Nevertheless, the jet formation mechanism can be the same in all astrophysical sources, with black hole mass and spin as the main driving parameters \citep{Foschini14}. From this point of view, LLCs\footnote{From now on, unless specified otherwise, the LLCs acronym will refer only to high-excitation sources.} have something in common with the other class of AGN known as narrow-line Seyfert 1 galaxies (NLS1s). NLS1s are characterised by a small full-width at half maximum (FWHM) of their permitted lines (FWHM$< 2000$\kms\ by definition, \citealp{Osterbrock85, Goodrich89}). The physical interpretation for this observational property is that the low FWHM corresponds to a low rotational velocity of the broad-line region gas surrounding a relatively less massive central black hole ($M_{\rm BH}$ $\approx$ 10$^6$-$10^8$ M$_\odot$, \citealp[e.g.,][]{Cracco16, Chen18, Dallabonta20}). In comparison, other types of AGN, such as broad-line Seyfert 1 galaxies, usually host black holes with higher masses ($M_{\rm BH}>10^{8}$ M$_\odot$). Since BLS1s and NLS1s should not be considered as a completely disjoint class of AGN \citep{Sulentic00, Laor00, Cracco16}, following the quasar main sequence scenario \citep{Marziani18b}, according to which there is a smooth transitions within the two classes, the difference in the size of the black hole mass could also be due to the H$\beta$ line threshold used to define each class \citep{Paliya24}. Moreover, the  difference within the two classes could also reside in the shape of the H$\beta$ line profile \citep{Cracco16}, Lorentzian profile for NLS1s and Gaussian profile for BLS1s instead. While the former corresponds to the turbulent motion of the clouds, the latter can be associated with Keplerian motion of the line-emitting clouds, and this can be explained by the different geometry and/or dynamics of the BLR for the two categories of Seyfert galaxies \citep{Kollatschny11}.\\
Regarding the radio emission, some NLS1s do harbour relativistic jets of relatively low power, as expected from their low-mass black holes \citep{Foschini15}. In radio, their spectrum can be as flat as in classical blazars ($\alpha_{\nu} \leq 0.5$, where F$_{\nu} \propto \nu^{-\alpha_{\nu}}$), indicating that their jets are oriented toward us, or steep ($\alpha_{\nu} > 0.5$), suggesting that the relativistic jet is misaligned \citep{Berton15a}.
Radio variability can affect the measurements of the radio spectral index in a short period, leading to mis-identification of the jetted sources as PSs \citep{Torniainen05}. Long-term radio observations are necessary to precisely constraint the value of the spectral index and, even in a single frequency, can help weed out the jetted sources from PS samples based on their variability.
Flat-spectrum NLS1s (F-NLS1) additionally constitute the low-luminosity tail of the radio luminosity function of flat-spectrum radio quasars \citep{Berton16c}, suggesting that they could represent an early evolutionary stage of this class of blazars \citep{Berton17}. \par
Given their similarities, it was suggested that LLCs and jetted F-NLS1s may be the same class of objects seen from different orientations. LLCs are predominantly observed at high inclination, whereas NLS1s, similar to other Type 1 AGN, are mostly seen pole-on \citep{Antonucci93, Komossa06}. \citet{Berton16c} built the radio luminosity function at 1.4 GHz of a sample of LLCs from \citet{Kunert10a}. By adding to it a model of relativistic beaming \citep{Padovani92}, they concluded that LLCs could be F-NLS1s observed at high inclination which was confirmed by \citet{Berton18a}.  \par

However, a major issue for the model may come from the host galaxies of LLCs. Jetted NLS1s are overwhelmingly hosted by disc galaxies \citep{Anton08, Kotilainen16, Olguiniglesias17, Jarvela18, Berton19a, Olguiniglesias20, Vietri22, Varglund22}, as expected from AGN with relatively low-mass black holes. On the contrary, the host galaxies of the general PS population tend to be large, bright ellipticals \citep{Devries98, Snellen98, Devries00, Labiano07, Kosmaczewski20}. If LLCs share the same host galaxy characteristics as the PS population, it becomes evident that unification with NLS1s is not feasible. However, some exceptions do exist, with PS disc hosts reported in the literature \citep{Orienti10a, Johnston10, Morganti11}. In light of these considerations, we decided to study the host galaxy morphology of the sample of LLCs used in the luminosity function study by \citet{Berton16c}. Our goal is to verify whether LLCs live in elliptical galaxies, like the majority of PS, or if they are hosted by disc galaxies as NLS1s.\\
In Sect.~\ref{sec:sample} we describe the sample selection and the data analysis method we used, while in Sect.~\ref{sec:sources} we present the results for single sources. In Sect.~\ref{sec:discussion} we discuss the results and give conclusions. Sect.~\ref{sec:summary} is a summary of this work.
Throughout this work, we adopted a standard $\Lambda$CDM cosmology with a Hubble constant $H_{0} = 67.8$ km s$^{-1}$ Mpc$^{-1}$, considering a flat Universe with the matter density parameter $\Sigma_{M} = 0.308$ and the vacuum density parameter $\Sigma_{vac} = 0.692$ \citep{Planck16}.

\section{Sample selection and data analysis}
\label{sec:sample}

\renewcommand{\arraystretch}{1.5}
\begin{table*}[ht!]
\caption[]{Sample sources from \citet{Berton16c}.}
\centering
\begin{tabular}{l l l l l l l l} \hline
SDSS Name                 &  RA          & Dec           & \textit{z}   & Scale & $M_{\rm BH}>$ & logL$_{1.4}$ & AGN type\\ 
                          &  (J2000.0)   & (J2000.0)     &     &(kpc/")& (M$_{\odot}$) & W Hz$^{-1}$ & \\ \hline
SDSS J002833.42+005510.9  & 00:28:33.423 &  +00:55:10.98 & 0.10429 & 1.975 & 8.94 & 24.77 & 2\\
SDSS J075756.71+395936.0  & 07:57:56.693 & +39:59:36.00 & 0.06578 & 1.303 & 7.13 & 24.00 & 2\\
SDSS J084856.57+013647.8  & 08:48:56.574 & +01:36:47.82 & 0.34987 & 5.088 & 7.05 & 25.54 & Int \\
SDSS J092607.99+074526.6  & 09:26:08.003 & +07:45:26.64 & 0.44169 & 5.868 & 7.28 & 26.05 & Int\\
SDSS J094525.90+352103.5  & 09:45:25.888 & +35:21:03.48 & 0.20777 & 3.505 & 7.23 & 25.24 & Int\\
SDSS J114311.01+053516.1  & 11:43:11.030 & +05:35:16.09 & 0.49694 & 6.264 & 8.84 & 26.26 & Int\\
SDSS J115727.61+431806.3  & 11:57:27.605 & +43:18:06.33 & 0.23012 & 3.790 & 7.68 & 25.58 & Int\\
SDSS J140416.35+411748.7  & 14:04:16.366 & +41:17:48.83 & 0.36042 & 5.187 & 7.96 & 25.96 & 2\\
SDSS J140942.44+360415.8  & 14:09:42.461 & +36:04:15.98 & 0.14864 & 2.676 & 8.24 & 24.91 & Int \\ 
SDSS J164311.34+315618.4  & 16:43:11.343 & +31:56:18.40 & 0.58668 & 6.808 & 7.44 & 26.20 & Int\\ \hline

\end{tabular}
\tablefoot{Columns: (1) Complete name; (2) Right ascension in h:m:s ; (3) Declination in d:m:s; (4) Redshift; (5) Scale calculated considering the Hubble constant to $67.8 $ km s$^{-1}$ Mpc$^{-1}$, a flat Universe with the matter density parameter $\Sigma_{M} = 0.308$ and the vacuum density parameter $\Sigma_{vac} = 0.692$; (6) Logarithm of the black
hole mass in M$_{\odot}$ from \citet{Berton16c}; (7) Log of the radio luminosity at 1.4 GHz from \citet{Kunert10c}; (8)  AGN type visually determined from the SDSS spectra: Type 2 or Intermediate (Int).}
\label{tab:sample}
\end{table*}

The LLCs sample we chose for this analysis is the same that was tested by \citet{Berton16c} as the parent population of the F-NLS1 sample and was used to build the radio luminosity function. In turn, this LLCs sample was derived from the study conducted by \citet{Kunert10a} on 44 LLCs, which show a steep radio spectral index with $\alpha_{\nu} > 0.7$ between 1.4 and 4.85 GHz. The steepness of the radio spectral index was verified based on multiple data available, like Low Frequency Array, Westerbork Northern Sky Survey, Texas Survey of Radio Sources at 365 MHz, Faint Images of the Radio Sky at Twenty-centimeters, 87GB and Effelsberg (single dish) measurements.
\citet{Kunert10b} cross-matched this 44 LLCs sample with the Sloan Digital Sky Survey (SDSS) DR7 spectroscopic archive, identifying 29 sources between HERG, LERG, and unclassified objects at $z <$ 0.9. On these 29 sources, \citet{Berton16c} applied the same completeness criterion used to select F-NLS1s (limit in magnitude of $i <$ 19.1 and in redshift of $z <$0.6, following \citet{Richards02}), resulting in a final sample of ten HE LLCs.\\
This final sample is shown in Tab.~\ref{tab:sample}, reporting also the main physical parameters characterising the sources.
As discussed in \citet{Berton16c}, the distribution of the black hole masses of LLCs is similar to the one of F-NLS1s, showing a median value between $10^{7.5}$ and $10^8$ M$_{\odot}$ in both samples (Tab.~\ref{tab:sample}). The two black hole mass distributions have been compared with a Kolmogorov-Smirnov test, confirming that they can be drawn from the same population. The distribution of the luminosity functions for LLCs and F-NLS1s sample, based on the measurements of the radio luminosity at 1.4 GHz, shows a similar scenario. The only difference between the two distributions lies in the slope of the luminosity function, being steeper for the LLCs to respect the F-NLS1s. 
According to our knowledge, the biggest sample of NLS1s with enhanced kpc-scale radio emission is the one studied by
\citet{Singh18}. The authors of this paper analysed the kpc-scale radio properties of almost five hundred of optically selected NLS1s, using several radio catalogues, aiming to study the nature of their radio jets. Within their sample of NLS1s, they found both steep and flat radio spectra, with 1.4 GHz radio luminosities of these sources span over a wide range, between $10^{22}$ and $10^{27}$ W Hz$^{-1}$. The range of radio luminosity at 1.4 GHz of our LLCs, going from $10^{24}$ to $10^{26}$ W Hz$^{-1}$, lies in the broader range found in \citep{Singh18}.

To determine the morphological type of the host galaxies, we used optical images obtained with the 3$\pi$ Steradian Survey of Pan-STARRS1 (PS1, \citet{Chambers16}) in five different filters (\textit{g}$_{P1}$, \textit{r}$_{P1}$, \textit{i}$_{P1}$, \textit{z}$_{P1}$, \textit{y}$_{P1}$). The stacked images from this survey have a mean 5$\sigma$ point-source sensitivity of 23.3, 23.2, 23.1, 22.3, and 21.4 mag (AB) and a median seeing of 1.''31, 1.''19, 1.''11, 1.''07, and 1.''02 for \textit{grizy}$_{P1}$ respectively. Furthermore, the median 50$\%$ completeness for the PS1 \textit{grizy}$_{P1}$ filters is 23.2, 23.2, 23.1, 22.3 and 21.2 respectively. From the second data release (DR2; \citet{Flewelling20}) of the PS1 3$\pi$ survey, we used stacked images, combined in each of the five bans. We downloaded the cutout images, each one centred on the NASA/IPAC Extragalactic Database position of the AGN. The size of each image, going from a minimum of 400 pixels to a maximum of 1000 pixels, was chosen based of the number of stars available in the field of view for point spread function (PSF) modelling.\\
When data in multiple filters are available, it is customary to minimise parameter degeneracy by extracting the galaxy structure from the band with the best host-to-AGN contrast ratio and resolution \citep{Matsuoka14}. Since the host galaxy should be brighter in the closest band to the IR, we choose the \textit{y}-band images as a reference. The \textit{y}$_{P1}$ filter, with  98$\%$ completeness, also offers the best coverage.
The host galaxy profile in the \textit{y}-band was modelled with a 2D-decomposition using the algorithm GALFIT version 3.0.5 \citep{Peng10}, which allows to fit simultaneously the multiple components that contribute to the sources' light profile. The procedure we followed is described in detail in \citet{Vietri22}. To quickly summarise, we proceeded as follows. The AGN emission must be modelled together with its host galaxy since it strongly contributes to the total profile. This approach is usually applied to Type 1 AGN observed at low inclination, for which the contribution from the active nucleus can overshine the one coming from the host galaxy. Since the AGN can be considered a point-like source, it is possible to model it with a stellar PSF. GALFIT can directly extract a PSF using profiles of isolated stars in the same field of view as the AGN and far from the image edges, modelled with several Sérsic functions. In the case of Type 2 AGN, observed at higher inclination instead, a PSF could no longer be needed to model the AGN contribution. In general, when a single PSF is not enough to reproduce the whole profile, the host galaxy is modelled adding Sérsic functions until a satisfactory result is obtained in terms of reduced chi-square $\chi_{\nu}^2$. \\
For this work, we proceeded as follows: first of all, we built a PSF for the y-band image of every source, and then we checked if a PSF was needed (depending on the AGN type) and if it was enough to model the total light profile. If it was not, we added a Sérsic function, a single one was enough at this level of resolution, to model in turn, the bulge or the disc component. 
Then, we measured a mean value of the background near each of the targets and fixed its value in the fitting procedure. We also estimated $\sigma_{sky}$ by measuring the standard deviation of the sky values in several empty regions of the image and taking the average of those. The $\sigma_{sky}$ was used to estimate the errors of the parameters related to the GALFIT procedure. To verify the reliability of the model produced by GALFIT, and that the output parameters were independent of the initial values fed to the algorithm, we also repeated the procedure by changing the input parameters \citep{Dewsnap23}. The details of each model are reported in the next section. 

Moreover, variations in stellar population and the dust attenuation can affect the morphology and the structural parameters of a galaxy, and let them vary with the wavelength. For example, in disc-like galaxies, the bluer bands will be dominated by the emission from the disc, while in the redder bands, the emission from the bulge becomes the dominant component \citep{Haussler13, Zhuang22}. This difference can be reflected in the value of the Sérsic index, which \citet{Kelvin12} demonstrated to be lower for the smaller wavelength and higher for the longer ones, showing an increase with the wavelengths itself. Due to this dependence of the morphology on the wavelengths, we decided to conduct an additional multi-filter analysis to compare (and possibly confirm) the results between different filters. We particularly aimed to verify whether the Sérsic index remains constant in different bands. To our knowledge, only a few morphological studies of the AGN host galaxy have been conducted to date by taking into account the wavelength dependency in the analysis (\cite{Martorano23}, \cite{Zhuang22}).\\
For this purpose, we used an extended version of GALFIT, GALFITM \citep{Haussler13, Vika13}, to perform the 2D photometric decomposition of the light profile simultaneously in different filters.
GALFITM is capable of processing a flexible number of images of the same sky region in various bands by applying a single, wavelength-dependent model to these images. Similarly to GALFIT, this model can consist of one or more component functions, each of them with several parameters. The multi-wavelength approach involves replacing the free parameters with functions of wavelength, such as Chebyshev polynomials, and replacing the standard parameters with the coefficients of these polynomials. The user can also specify the maximum order of each polynomial function, choosing if the standard parameter has to be constant with the wavelength with a zero-order Chebyshev polynomial, for example, or if it has to vary as a constant function of the wavelength, or giving the function the freedom to interpolate the data \citep{Haussler13, Vika13}.

We chose to use GALFITM on the bluest and reddest bands (\textit{g}$_{P1}$ and \textit{y}$_{P1}$, respectively), since the Sérsic index has been observed to vary at the shorter and longer wavelengths \citep{Kelvin12}. We also extended the analysis to a central band between the reddest and the bluest ones, chosen as the one containing H$\alpha$ in each source, as a tracer of star formation activity, which can affect the morphology of the galaxy, as explained before.
We remark that our main goal was to understand whether the host galaxies were early- or late-type. From this point of view, the Sérsic index, $n$, of the galactic bulge is an important indicator. For this reason, we also used a multiwavelength approach to better constrain the value of the Sérsic index. Small values of $n$, typically less than two, indicate the presence of a pseudobulge and are associated with late-type morphologies and other features such as bars, spiral arms, or rings. Classical bulges, on the other hand, are normally found in early-type galaxies, and show a higher Sérsic index ($n \geq$ 4). 

\section{Single sources analysis}
\label{sec:sources}
As mentioned in the Introduction, since LLCs are typically observed at a high inclination, they should predominantly be Type 2 AGN. Consequently, the PSF may no longer be necessary in the model to fit the host galaxy profile. Our sample of LLCs comprises Type 2 AGN and Intermediate type AGN \citep{Berton16c}. For three sources, the PSF was not indeed needed to model the host galaxy. In three particular cases, instead, only a PSF is needed to model the brightness profile. For the remaining sources, a combination of PSFs and Sérsic functions has been used to properly model the host profile. The sky value was held constant throughout all the fitting procedures. Some sources exhibit a nearby companion within the fitting region, and in these cases, a PSF model was employed to simultaneously fit the objects close to our target.
The morphological classification of each object was also confirmed by varying the input parameters, and GALFIT consistently converged to the same result. The multi-band analysis yielded results consistent with those obtained from the single-band analysis, except in one case, which will be discussed later. Further details of the GALFITM analysis will be described in the following subsections.\\
To estimate all parameter errors related to the GALFIT procedure, we fit the source by adding and then subtracting $\sigma_{sky}$ from the sky value. For each parameter, the positive and negative errors were determined as the differences between the resulting values and the best-fit values, respectively.
Furthermore, we searched the literature to determine whether our sources have been previously analysed in other morphological studies for comparison and possible confirmation of our results.
We found that most of the LLCs in our sample have been previously classified by \citet{Nascimento22}. 
The authors only made a visual classification of a sample of CSS and GPS sources using Sloan Digital Sky Survey (SDSS) images. Three distinct classifiers worked in this sense, and the morphological class was established when at least two of them agreed. For this reason, we need to consider that visual analysis can also bring misleading classification. Only one source was found to be analysed also by \citet{Urbano18}, which conducted a morphological study of hosts Type 2 AGN with images from the \textit{Hubble Space Telescope} (\textrm{HST}), using GALFIT.
In the next subsections, we will present the analysis procedure followed and the results obtained for each source individually.

\subsection{SDSS J002833.42+005510.9} 
\begin{figure}
    \includegraphics[width=9cm]{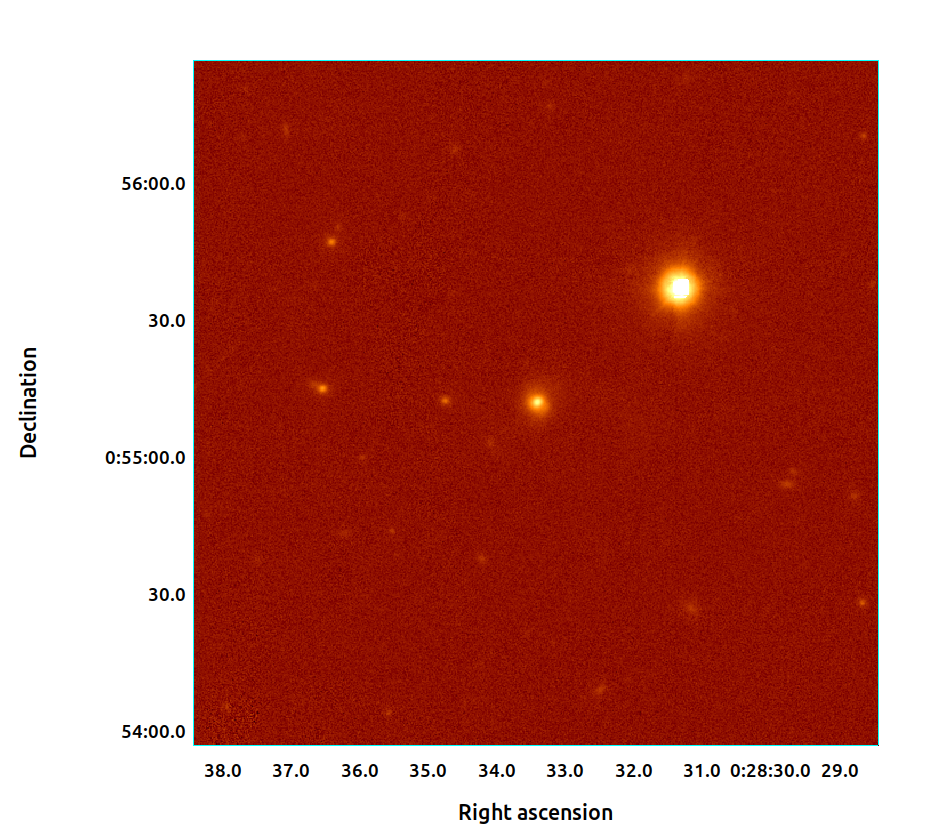}
    \caption{PS1 i-band image of J002833.42+005510.9.}
    \label{fig:image1}
\end{figure}

\begin{figure}
    \includegraphics[width=9cm]{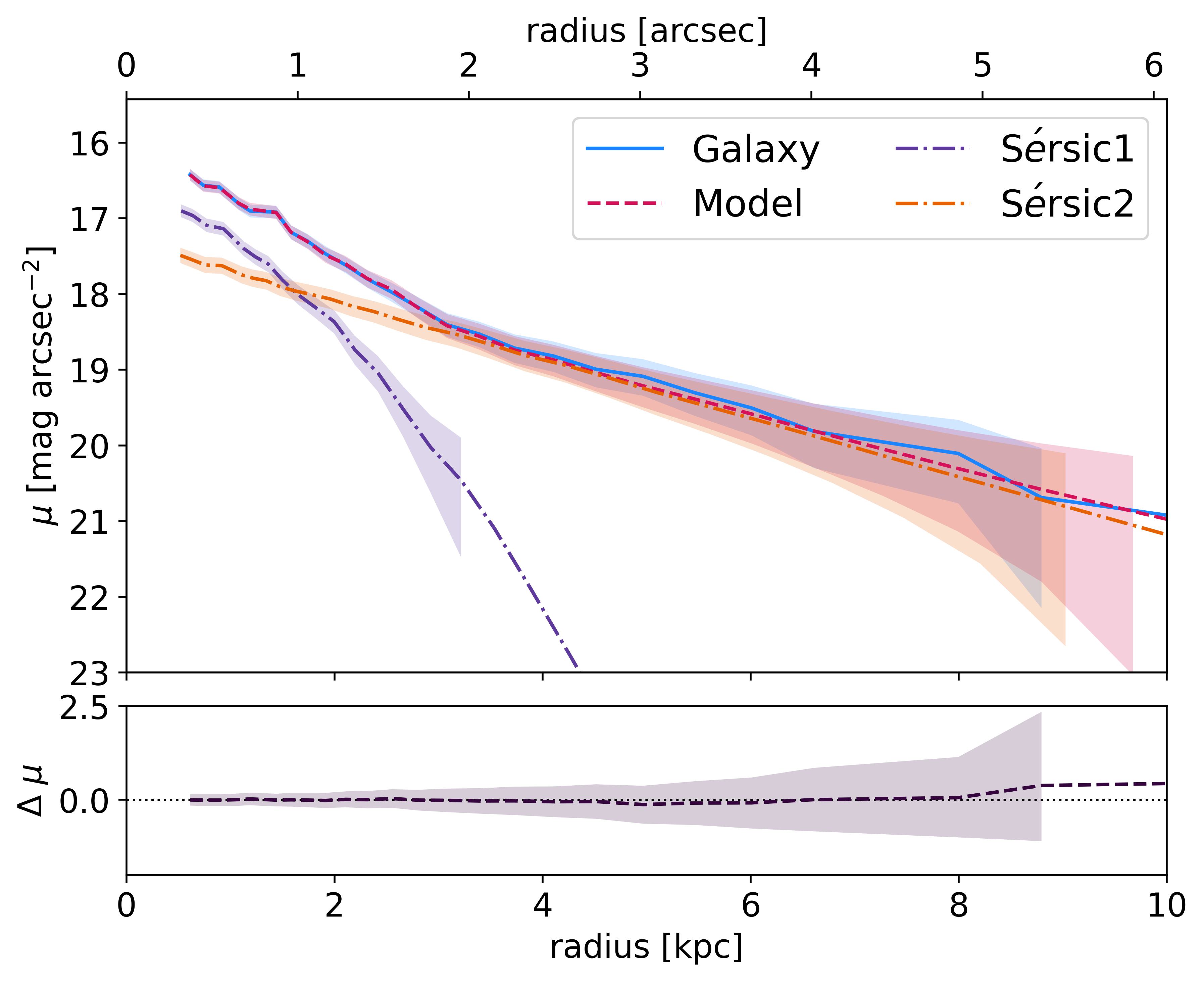}
    \caption{Radial surface brightness profiles of J0028+0055: the observed profile and the model. Legend in the plot. The shaded area around each profile describes the associated errors. The low panel shows the variation in magnitude.}
    \label{fig:host1}
\end{figure}

We performed the host galaxy photometric decomposition of y-band image of J002833.42+005510.9 <4C +00.03>, ($z$ = 0.104, Type 2 AGN), to determine its morphological type. As explained in the previous section, since this is a Type 2 AGN, a PSF model was not needed to remove the AGN contamination. The best fit for this galaxy was achieved with two Sérsic functions, one for the bulge and one for the disc (S1 and S2, respectively). The parameters of the best fit, related to the y-band images, for all the sources, are listed in Tab.~\ref{tab:tab1}. Although the bulge is only marginally resolved, as shown by the very small effective radius $R_{\rm e}$, the low bulge Sérsic index indicates the presence of a pseudo-bulge. From this analysis, we can affirm that this LLC source is hosted in a late-type galaxy. The negative errors, related to several parameters, appear to be very high, especially for the $R_{\rm e}$ of the disc. The reason could lie in GALFIT itself, which tries to fit the sky instead of the galaxy when the fixed value of the sky parameter is very low (as in the case of the \textit{sky} - \textit{$\sigma_{sky}$} fit). This can also explain why it happens only with the disc component. We adopt the definition "not applicable" \textit{(N/A)} in case of high errors.
For this source, we applied GALFITM to the $g_{P1}$, $i_{P1}$ (containing H$\alpha$) and $y_{P1}$ bands. We left the total magnitude, the effective radius $R_{\rm e}$ and the Sérsic index free to vary for both Sérsic functions, and fit them with a 2nd-order Chebyshev polynomial to give the freedom to interpolate data precisely. We found consistent results for S1 and S2 to the one obtained for the y-band, confirming that this AGN is hosted in a late-type galaxy with a pseudo-bulge. 
From the literature, we found that this galaxy has been visually classified as a spiral by \citet{Nascimento22}.
Since the galaxy features should be more prominent in the optical band, we decided to show all i-band image from the PS1 for each source. The i-band image of this source is shown in Fig.~\ref{fig:image1} where it is possible to notice that this object is surrounded by a diffuse emission that can resemble a disc. 
The radial surface brightness profile is shown in Fig.~\ref{fig:host1}, where it is possible to notice that the galaxy profile is well represented by the model made up of the bulge component at smaller radii and the disc component at larger radii. 

\renewcommand{\arraystretch}{1.5}
\begin{table*}[ht!]
\caption[]{Best fit parameters of the ten LLC sources.}
\centering
\begin{tabular}{l l l l l l l l l }
\hline\hline
Source & Function   &  mag                            & $R_{\rm e}$   (kpc)                      & $n$                            & axial  ratio                    & PA (\textdegree)   & notes & $\chi^{2}_{\nu}$ \\ \hline
J0028+0055 & S1 & 16.29 $\substack{+0.39\\-1.44}$ &  1.35 $\substack{+0.04\\-1.24}$ & 0.60 $\substack{+0.02\\-0.58}$ & 0.85 $\substack{+0.06\\-0.14}$ & -60.57 $\substack{+4.74\\-10.76}$ & bulge & 1.057$_{-0.052}^{+0.701}$\\
& S2 & 14.89 $\substack{+0.68\\-4.03}$ &  4.49 $\substack{+0.60\\\textit{N/A}}$  & 1.02 $\substack{+0.65\\-0.16}$ & 0.88 $\substack{+0.03\\-0.17}$ & 34.55 $\substack{+4.96\\ \textit{N/A}}$ & disc &  \\ \hline
J0757+3959 & S & 13.89 $\substack{+0.05\\-1.30}$ & 2.02 $\substack{+0.12\\-2.42}$ & 2.72 $\substack{+0.62\\-1.00}$ & 0.85 $\substack{+0.02\\-0.05}$ & 3.34 $\substack{+3.01\\-2.41}$ & bulge & 1.004 $_{-0.001}^{+0.955}$\\
& exp disc & 13.37 $\substack{+0.90\\-3.27}$ & 4.61 $\substack{+1.73\\ \textit{N/A}}$  & 1 & 0.86 $\substack{+0.06\\-0.14}$ & -44.38 $\substack{+0.77\\ \textit{N/A}}$ & disc & \\ \hline
J0848+0136 & S & 16.09 $\substack{+1.67\\\textit{N/A}}$ &  14.21 $\substack{+10.68\\\textit{N/A}}$ & 1.79 $\substack{+1.14\\\textit{N/A}}$ & 0.76 $\substack{+0.18\\-0.01}$ &  -66.69 $\substack{+10.27\\-9.48}$ & disc & 1.106 $_{-0.056}^{+0.753}$ \\  \hline
J0926+0745 & PSF & 15.77 $\substack{+0.00\\-0.00}$ & & & & & AGN & 1.179 $\substack{+0.000\\-0.000}$ \\ \hline 
J0945+3521 & PSF & 16.67 $\substack{+0.16\\-0.53}$ &  & &  &  & AGN & 1.093 $_{-0.113}^{+0.779}$\\
& S1 & 15.37 $\substack{+0.82\\-5.21}$ & 5.24 $\substack{+2.19\\\textit{N/A}}$  & 1.96 $\substack{+1.49\\-1.34}$ & 0.91 $\substack{+0.00\\-0.05}$ & -15.31 $\substack{+13.91\\-1.46}$ & bulge/disc & \\ \hline
J1143+0535 & PSF & 16.97$_{-0.05}^{+0.06}$ & & & & & AGN & 1.136$_{-1.125}^{+1.988}$ \\ \hline
J1157+4318 & PSF & 16.23 $\substack{ \textit{N/A} \\-0.67}$ &  & &  &  & AGN & 1.108 $_{-0.090}^{ \textit{N/A} }$\\
& S1 & 15.43 $\substack{ \textit{N/A}\\-6.23}$ & 4.63 $\substack{ \textit{N/A} \\ \textit{N/A}}$  & 4.47 $\substack{ \textit{N/A} \\-1.60}$ &  0.56 $\substack{ \textit{N/A} \\-0.25}$ &  -5.87 $\substack{ \textit{N/A} \\ \textit{N/A}}$ & bulge & \\ \hline
J1404+4117 & PSF & 18.16 $\substack{+0.18\\-1.29}$ &  & &  &  & AGN & 1.138 $_{-1.136}^{+0.039}$\\
& S1 & 16.19 $\substack{+0.86\\-4.59}$ & 6.06 $\substack{+3.47\\ \textit{N/A}}$  & 2.26 $\substack{+1.85\\-2.00}$ &  0.88 $\substack{+0.00\\-0.23}$ &  -19.43 $\substack{+0.48\\ \textit{N/A}}$ & bulge/disc &  \\ \hline
J1409+3604 & PSF & 17.04 $\substack{+0.21\\-0.75}$ &  & &  &  & AGN & 0.786 $_{-0.369}^{+0.002}$\\
& S1 & 15.08 $\substack{+0.20\\-3.18}$ & 5.89 $\substack{+2.83\\ \textit{N/A}}$  & 2.32 $\substack{+1.65\\ \textit{N/A}}$ &  0.66 $\substack{+0.08\\-0.01}$ &  0.66 $\substack{+6.98\\-4.07}$ & bulge/disc & \\ \hline
J1643+3156 & PSF & 16.59 $_{-0.12}^{+0.13}$ &  & &  &  & AGN & 1.151 $_{-1.023}^{+1.042}$ \\ \hline
\end{tabular}
\tablefoot{Columns: (1) short name (2) function used in the model: Sérsic function (S, if more than one S1 and S2), PSF, exponential disc (exp disc); (3) magnitude of the component in $y$-band; (4) effective radius; (5) S\'{e}rsic index; (6) axial ratio; (7) position angle; (8) physical interpretation; (9) Reduced chi-square.}
\label{tab:tab1}
\end{table*}

\subsection{SDSS J075756.71+395936.0}
\begin{figure}
    \includegraphics[width=9cm]{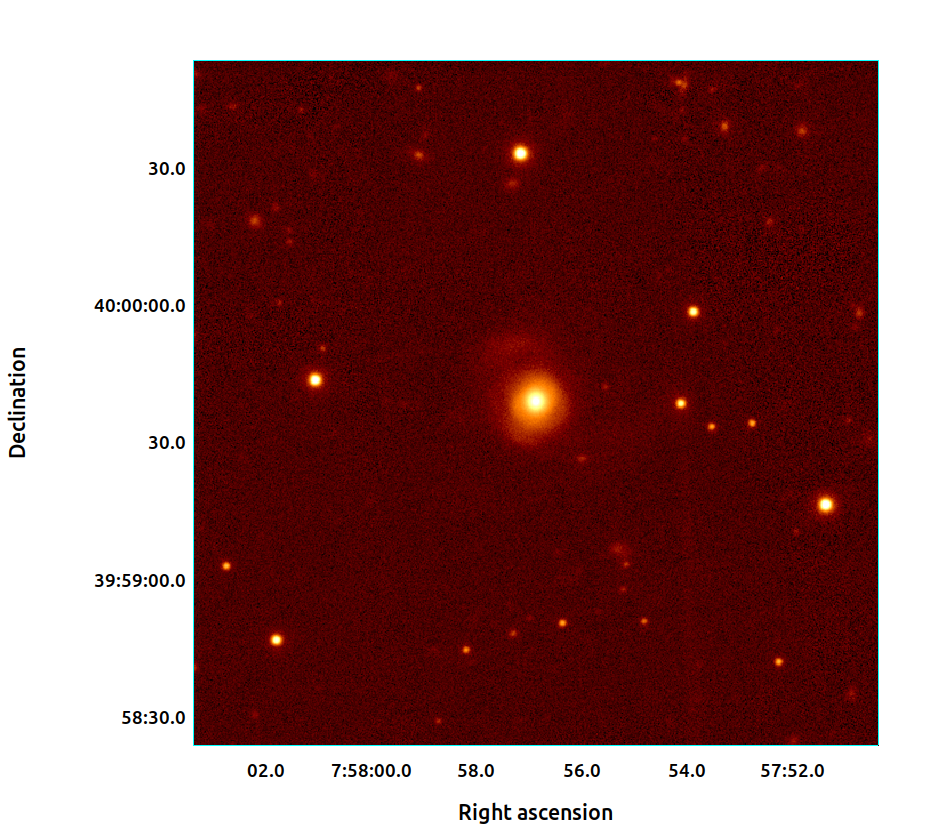}
    \caption{PS1 i-band image of J075756.71+395936.0.}
    \label{fig:image2}
\end{figure}

\begin{figure}
    \centering
    \includegraphics[width=9cm]{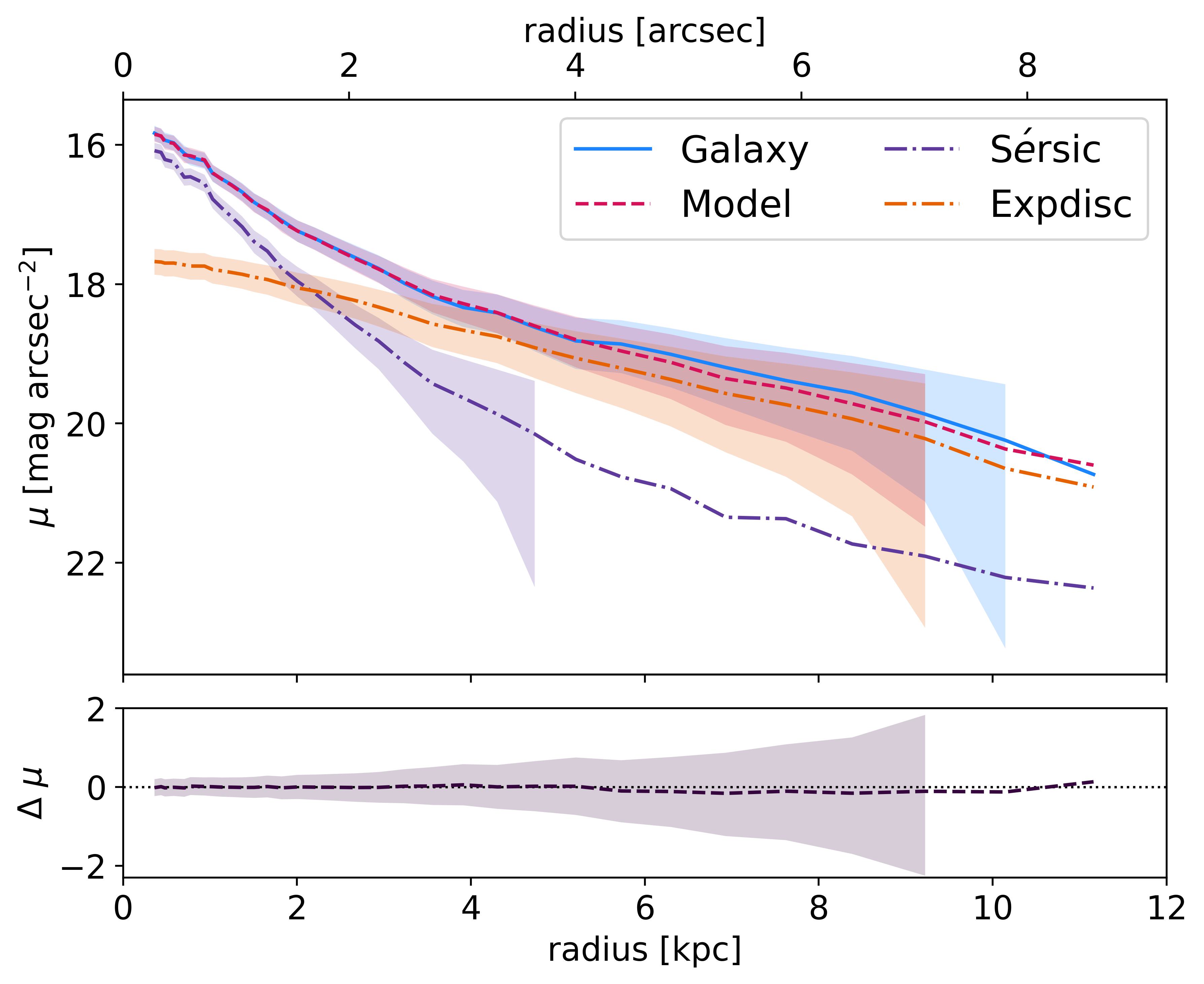}
    \caption{Radial surface brightness profiles of J0757+3959: the observed profile and the model. Legend in the plot. The shaded area around each profile describes the associated errors. The lower panel shows the variation in magnitude.}
    \label{fig:host2}
\end{figure}

In the case of J075756.71+395936.0 <B3 0754+401>, ($z$=0.065, Type 2 AGN), the PSF was not needed to properly model the source.
The best fit was achieved with a Sérsic function for the bulge, plus an exponential disc (parameters listed in Tab.~\ref{tab:tab1}). The Sérsic index of the bulge, being around 2, confirms the presence of a pseudo-bulge.
We used GALFITM with the $g_{P1}$, $i_{P1}$ (containing H$\alpha$) and $y_{P1}$ band. Again, we left the total magnitude, the effective radius R$_{\rm e}$ and the Sérsic index free to vary, for both Sérsic functions. We found consistent results for the bulge and the exponential disc with the one obtained for the y-band, confirming that this AGN is hosted in a late-type galaxy with a pseudo-bulge. 
The galaxy has been visually classified as a spiral by \citet{Nascimento22}.
The i-band image of this source is shown in Fig.~\ref{fig:image2} where the galaxy can be clearly identified as a disc galaxy, showing a prominent disc and outer halo, with another external feature that may be the relic of a past merger. The radial surface brightness profile is shown in Fig.~\ref{fig:host2}. The model deviates a bit from the galaxy profile at large radii, maybe due to the presence of the outer halo that has not been included in the model.

\subsection{SDSS J084856.57+013647.8}
\begin{figure}
    \includegraphics[width=9cm]{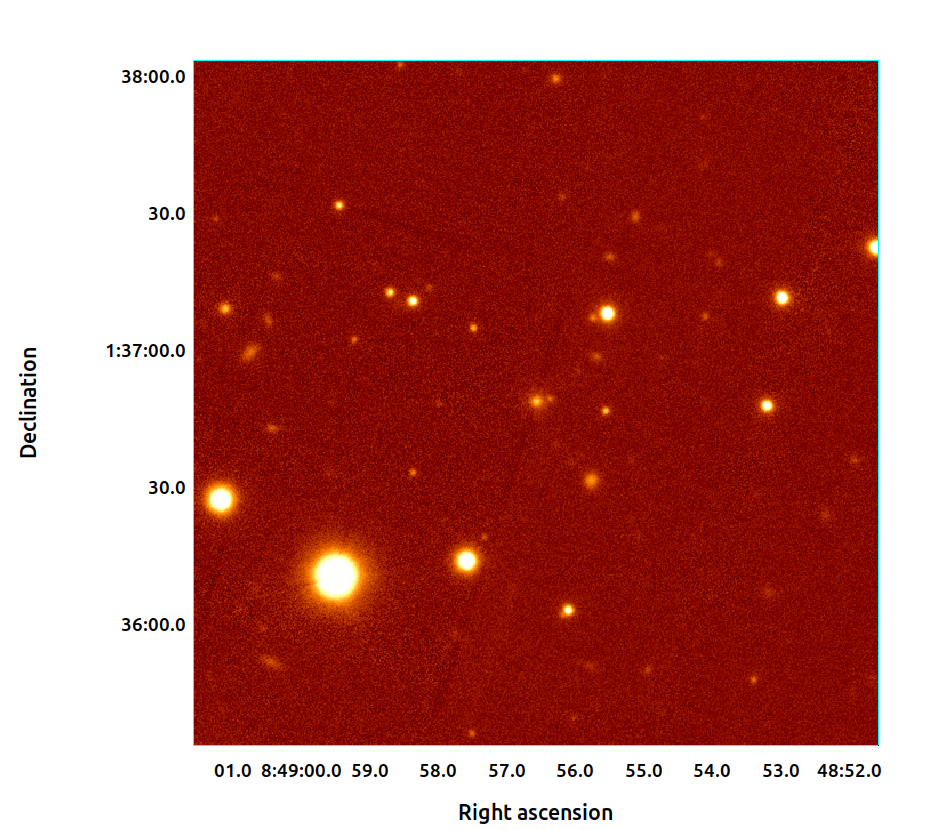}
    \caption{PS1 i-band image of J084856.57+013647.8.}
    \label{fig:image3}
\end{figure}

\begin{figure}
    \centering
    \includegraphics[width=9cm]{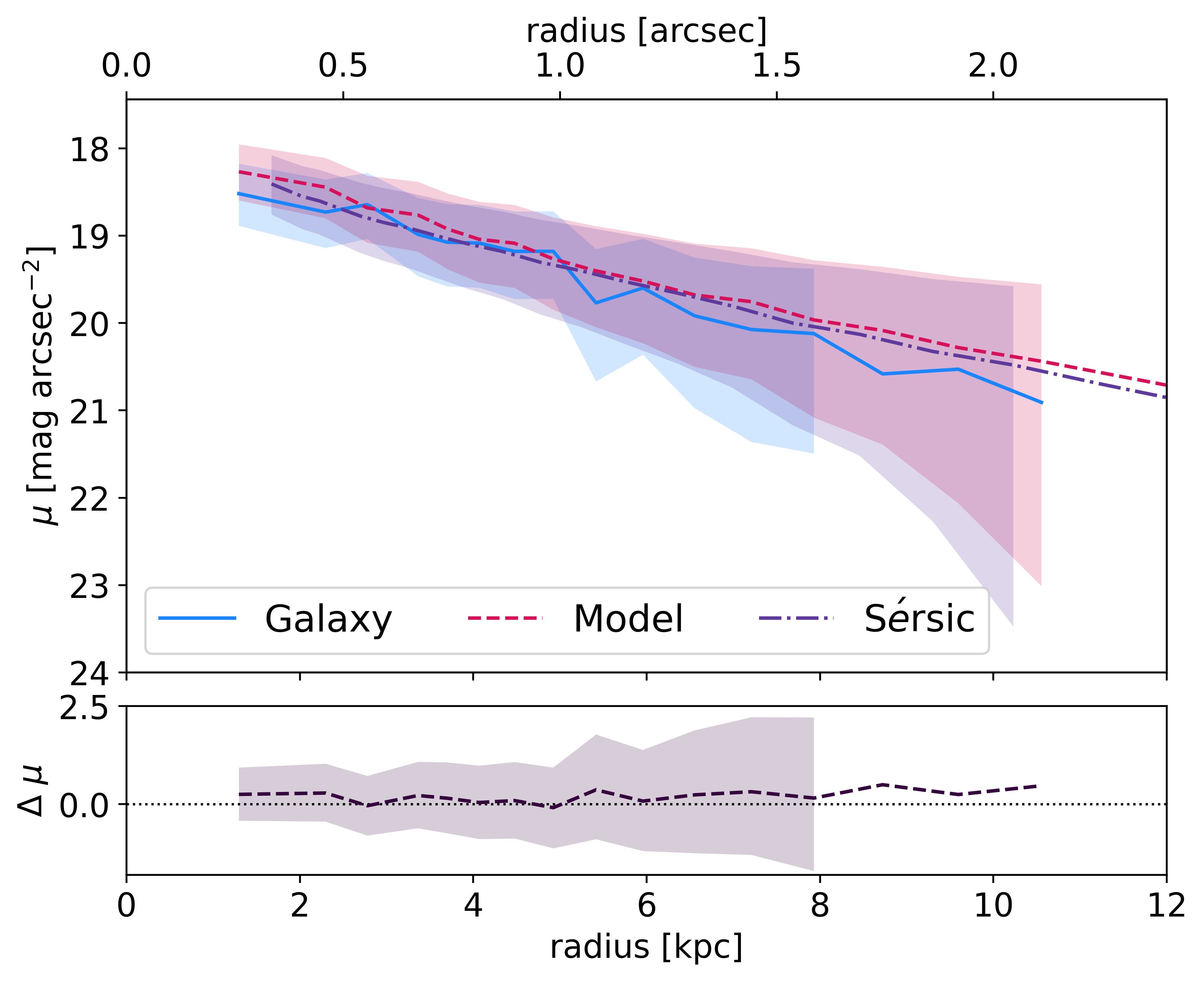}
    \caption{Radial surface brightness profiles of J0848+0136: the observed profile and the model. Legend in the plot. The discrepancy between the galaxy profile and the model could be due to the presence of another source very close to the target. The shaded area around each profile describes the associated errors. The lower panel shows the variation in magnitude.}
    \label{fig:host3}
\end{figure}

The third object, J084856.57+013647.8 (an Intermediate AGN, $z=$0.349), shows a possible second source very close to the target, as it is possible to notice in the i-band image shown in Fig.~\ref{fig:image3}. It is not straightforward to recognise this second feature as another object or as an external part of the outer halo or disc, due to the redshift and consequent high scale.
Only the source itself has been considered in the fit, and the best model was achieved with a single Sérsic function, without a PSF. Intermediate AGN, which are galaxies showing spectral properties that fall between Types 1 and Types 2, can be divided into several categories depending on the prominence of the broad component with respect to the narrow one. For this reason, AGN Type 1.9 or Type 1.8, for example, share several properties with Type 2, as a significant obscuration. This source can be classified as a Type 1.8 AGN, which explains the absence of the PSF in the fit. Parameters can be found in Tab.~\ref{tab:tab1}.
The Sérsic function exhibits a low Sérsic index, close to 1, and an high R$_e$, which can be identified with an outer disc or with GALFIT trying to fit the second source. This allows us to classify this host galaxy as a late-type one.
We also used GALFITM on the $g_{P1}$, $z_{P1}$ (which contains H$\alpha$) and $y_{P1}$ bands. We allow the total magnitude, the effective radius $R_{\rm e}$ and the Sérsic index to vary as a constant function of the wavelength, for both Sérsic functions. Although we obtained consistent results between the single- and the multi-band analysis, in the literature we found that \cite{Urbano18} measured a Sérsic index of 5 for the bulge of this galaxy, using \textrm{HST} data, indicating an elliptical galaxy.
In first instance, we thought that the discrepancy between the two different classifications could have been due to the different resolution of the used data and the presence of a second source very close to the target. To verify that, we smoothed the \textrm{HST} image of this source, obtained with the Advanced Camera for Surveys (ACS), by convolving it with a Gaussian kernel with a FWHM equal to the quadrature difference of the two former FWHM. The first thing we noticed from the \textrm{HST} smoothed image is a faint halo surrounding the galaxy and resembling the rest of a past merger.  We tried to decompose the light profile, using GALFIT, but it was not possible to obtain a reliable result. We started fitting the host galaxy with different Sérsic functions, without a PSF, and then we also tried with Gaussian and exponential disc functions. Each time we obtained a high effective radius $R_{\rm e}$ or disc scale length $R_{\rm s}$, meaning that it was trying to fit the outer halo within the model. Also, the $\chi^{2}_{\nu}$ was incredibly low, on the order of 0.01. When we took a look at the residual we understood why our attempts at modelling the galaxy were not successful: the galaxy showed a lopsided symmetry, that can be a sign of a past merger.  We also tried using Fourier modes to model this galaxy but we did not obtain a satisfying result in terms of $\chi^{2}_{\nu}$ and residuals.
It should be noted that \citet{Urbano18} did not report the values of the $\chi^{2}_{\nu}$ neither the residuals. Visually analysing the \textrm{HST} image and the residuals from the several attempts we made to model this source, we can affirm that it is almost impossible that this AGN is hosted in an elliptical galaxy. It is more reliable that this galaxy has gone through a recent merger which signs are clearly visible in the outer halo and its lopsided symmetry, and that it can be morphological classified as an irregular galaxy, rather than an elliptical.
Finally, the surface brightness profile, represented in Fig.~\ref{fig:host3}, shows that the model deviates slightly from the galaxy profile, probably due to the presence of the outer component that is not involved in the model.

\subsection{SDSS J092607.99+074526.6}
\begin{figure}
    \includegraphics[width=9cm]{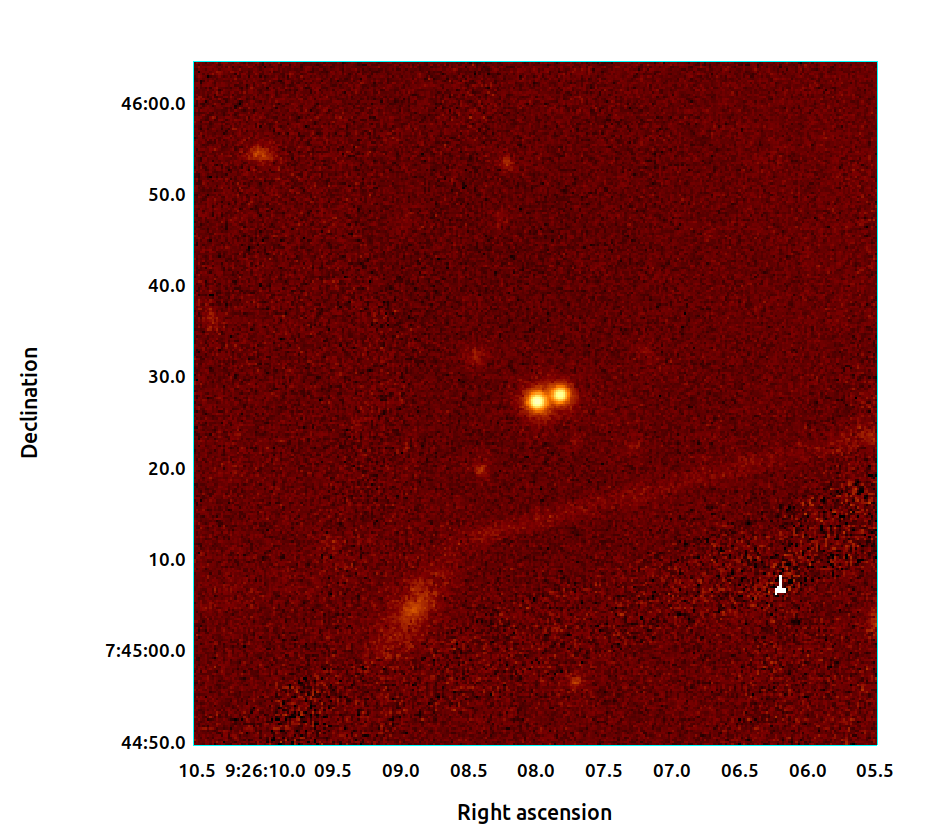}
    \caption{PS1 i-band image of J092607.99+074526.6.}
    \label{fig:image4}
\end{figure}

\begin{figure}
    \centering
    \includegraphics[width=9cm]{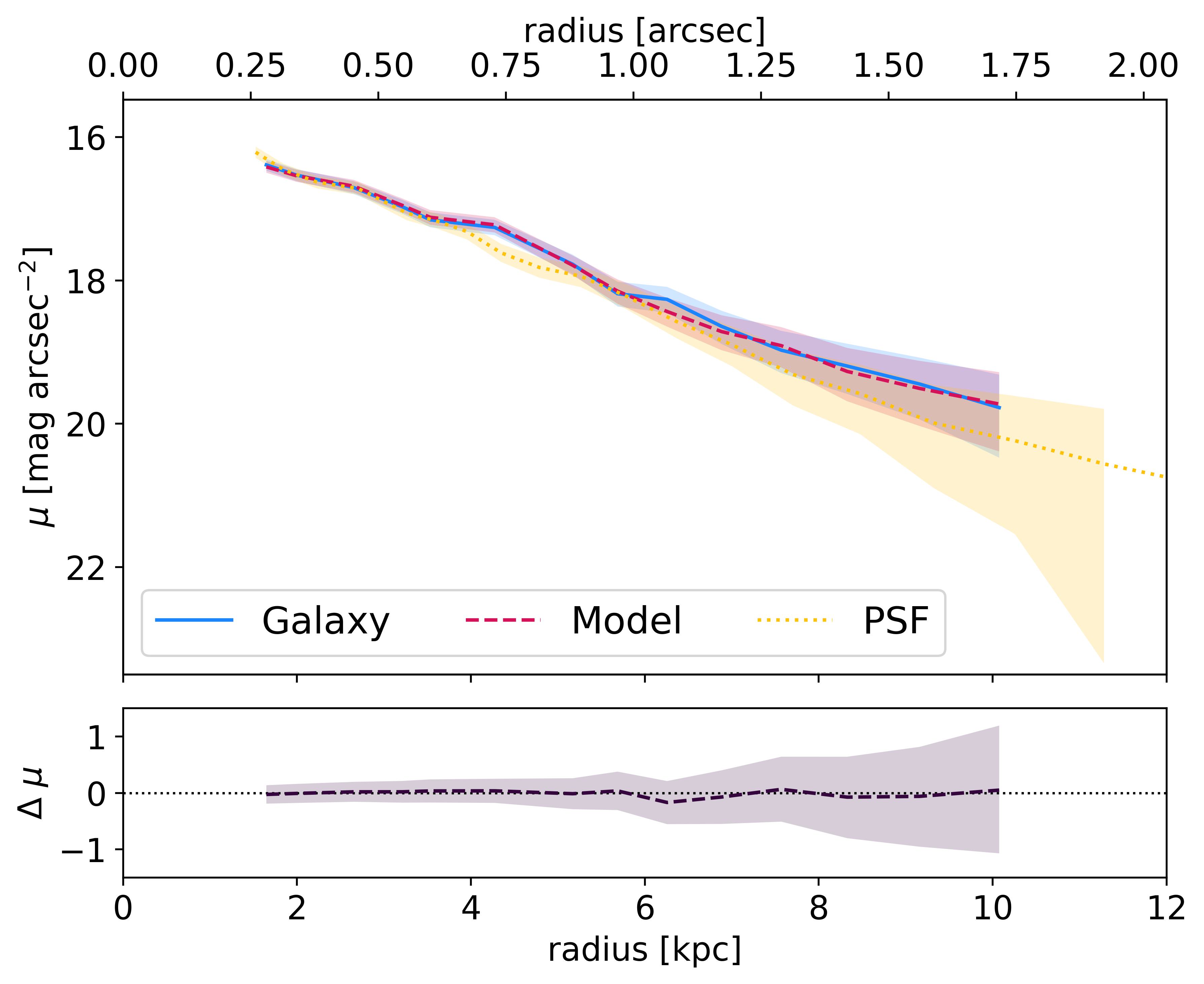}
    \caption{Radial surface brightness profiles of J0926+0745: the observed profile and the model. Legend in the plot. The shaded area around each profile describes the associated errors. The lower panel shows the variation in magnitude.}
    \label{fig:host4}
\end{figure}

Also in the case of J092607.99+074526.6 (Intermediate AGN, $z$=0.441), it seems that the galaxy has a companion, as shown in Fig.~\ref{fig:image4}. No results have been found in the literature on the classification of the companion. We decided to model the sources together with only 2 PSFs, each one for each source, and no more details were needed to obtain the best fit. 
The resulting reduced chi-square is $\chi^{2}_{\nu}$= 1.179$_{-0.00}^{+0.00}$. The PSF magnitude of the main source is PSF$_1$ mag= 15.77$_{-0.00}^{+0.00}$ while for the other source PSF$_2$ mag = 16.19$_{-0.00}^{+0.00}$. The errors related to the fit for the PSF magnitudes and the reduced chi-square are equal to zero.
Analysis made with GALFITM on the $g_{P1}$ and $y_{P1}$ band (containing H$\alpha$) confirms that only two PSFs are needed to properly model this system. Also, no results in the literature have been found about the morphological type of this peculiar object. For this reason, nothing can be said regarding the morphology of this galaxy. The surface brightness profile (Fig.~\ref{fig:host4}) shows good agreement between the galaxy profile and the model itself.

\subsection{SDSS J094525.90+352103.5}
\begin{figure}
    \includegraphics[width=9cm]{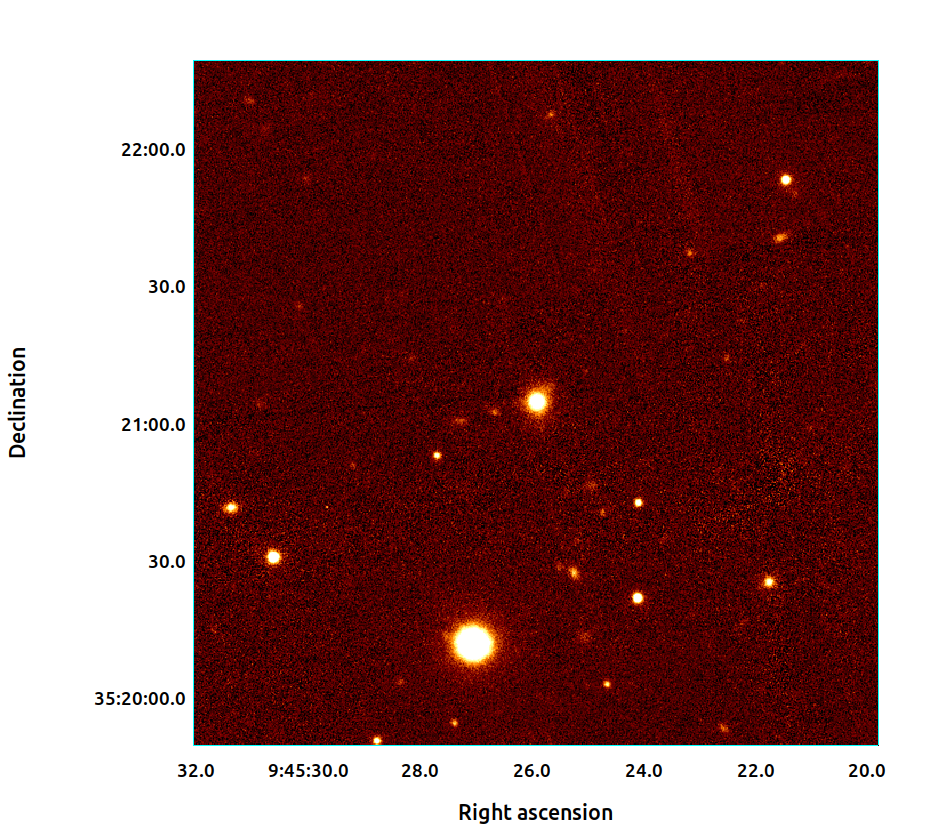}
    \caption{PS1 i-band image of J094525.90+352103.5.}
    \label{fig:image5}
\end{figure}

\begin{figure}
    \centering
    \includegraphics[width=9cm]{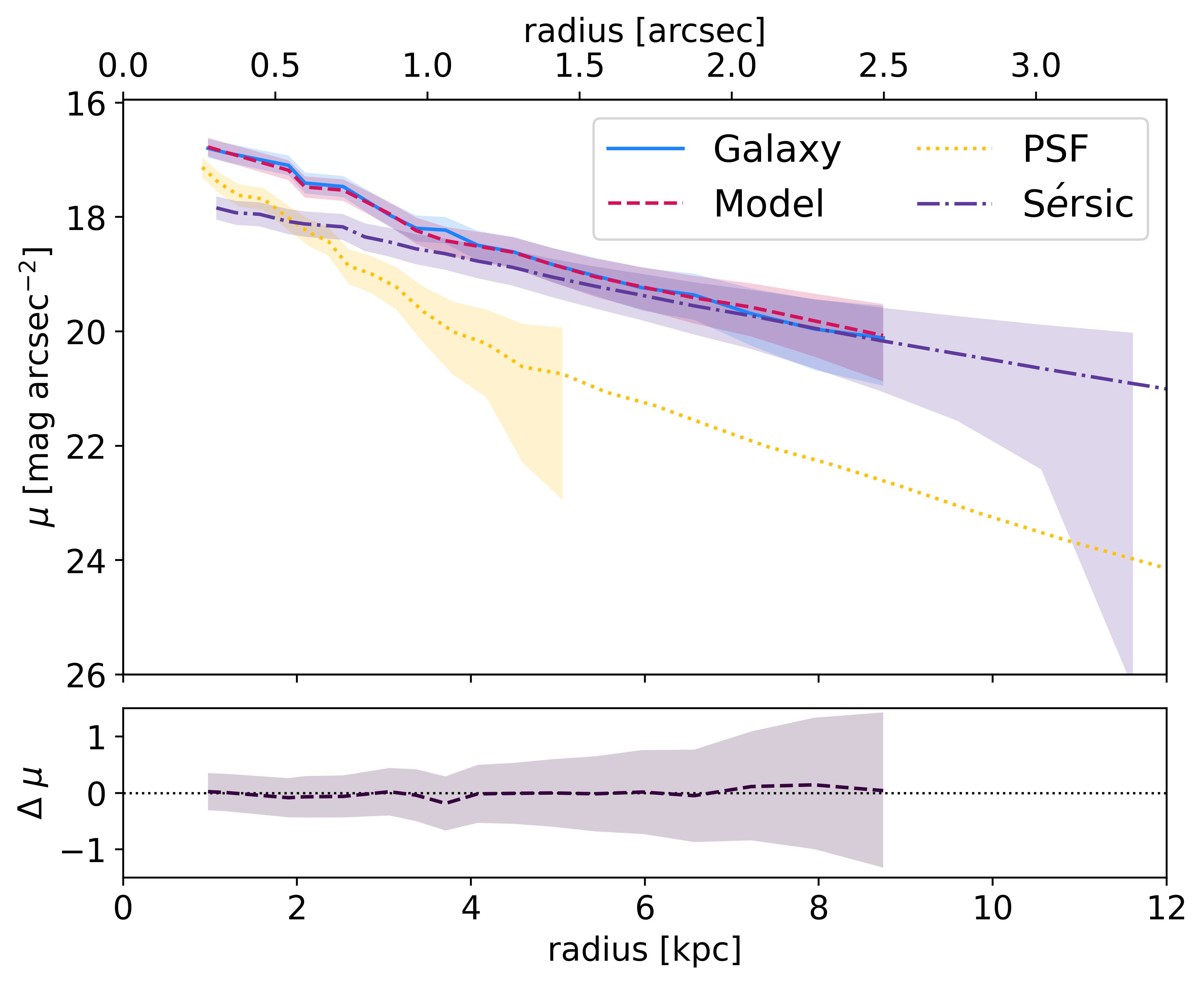}
    \caption{Radial surface brightness profiles of J0945+3521: the observed profile and the model. Legend in the plot. The shaded area around each profile describes the associated errors. The lower panel shows the variation in magnitude.}
    \label{fig:host5}
\end{figure}

For J094525.90+352103.5 <B2 0942+35A>, ($z=$0.207, Intermediate AGN), the PSF model was obtained with two Sérsic functions and the best galaxy fit was achieved with the PSF plus a Sérsic function. This source can be classified as Type 1.5 or 1.2, showing a grade of obscuration closer to Type 1 AGN than Type 2 and justifying the presence of the PSF in the fit. The best fit parameters can be found in Tab.~\ref{tab:tab1}. The low Sérsic index of the second component and the R$_e$ extending outside the bulge region could indicate that GALFIT is trying to fit something between a bulge and a disc. This can happen with late-type galaxy hosting pseudo-bulges when the Sérsic index is lower than two and so close to the Sérsic index of the disc. 
To confirm this result, we also used GALFITM with the $g_{P1}$, $i_{P1}$ (containing H$\alpha$) and $y_{P1}$ bands. We left the total magnitude free to vary, for both the PSF and the Sérsic function, while we allowed the effective radius $R_{\rm e}$ and the Sérsic index to vary as a constant function of the wavelength. We obtained consistent results for the single band analysis, confirming that this source is hosted in a late-type galaxy. This galaxy has also been visually classified as an irregular/merger by \citet{Nascimento22}. The i-band image (Fig.~\ref{fig:image5}) indeed shows a halo with irregular features surrounding the source.
The surface brightness profile (Fig.~\ref{fig:host5}) shows good agreement between the model and the galaxy profile.

\subsection{SDSS J114311.01+053516.1}
\begin{figure}
    \includegraphics[width=9cm]{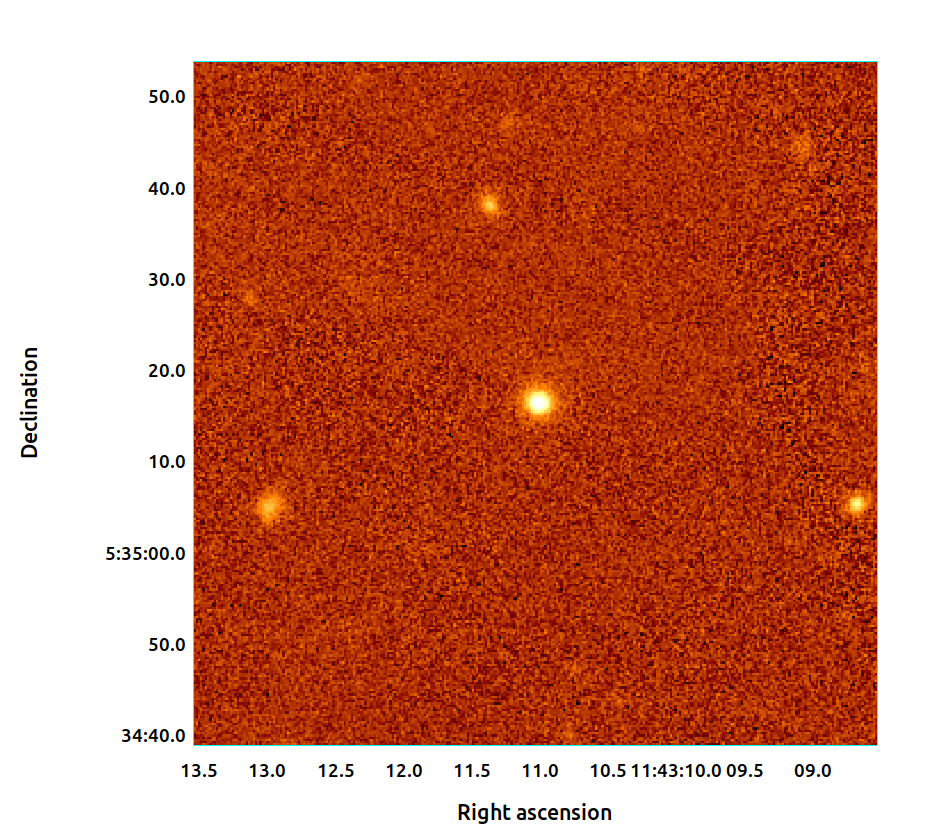}
    \caption{PS1 i-band image of J114311.01+053516.1.}
    \label{fig:image6}
\end{figure}

\begin{figure}
    \centering
    \includegraphics[width=9cm]{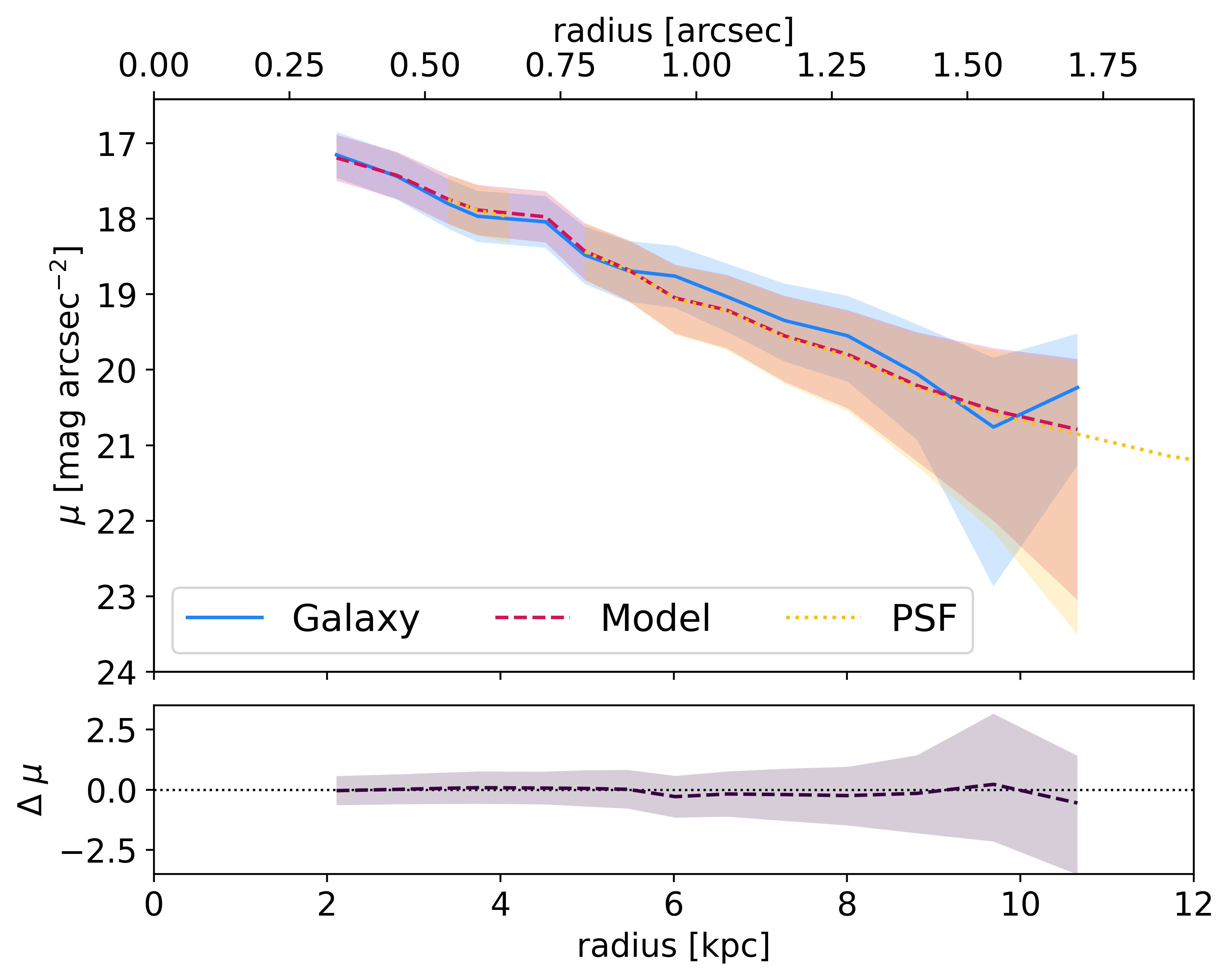}
    \caption{Radial surface brightness profiles of J1143+0535: the observed profile and the model. Legend in the plot. The shaded area around each profile describes the associated errors. The low panel shows the variation in magnitude.}
    \label{fig:host6}
\end{figure}

For SDSS J114311.01+053516.1 ($z=$ 0.496, Intermediate AGN), it was not possible to determine the host morphological type. Building a model with Sérsic functions only brought to non-physical parameters. The best fit for the y-band image, in this case, was achieved with a single PSF and a reduced $\chi^{2}_{\nu}$ of 1.136$_{-1.125}^{+1.988}$. The best PSF model was obtained with two Sérsic functions. The magnitude of the PSF component is $mag_{PSF}$ = 16.97$_{-0.05}^{+0.06}$. 
We used GALFITM on the $g_{P1}$ and $y_{P1}$ bands (containing H$\alpha$), leaving the magnitude of the PSF free to vary, which confirms that a single PSF was sufficient to fit the light profile of this source. In this case, nothing can be said about the morphology of this object. This source has been visually classified as irregular/merger by \citet{Nascimento22}, while the i-band image (Fig.~\ref{fig:image6}) does not show particular features. The radial surface brightness profile (Fig.~\ref{fig:host6}) shows that the model is quite in agreement with the galaxy profile, but it deviates a bit from it at large radii.

\subsection{SDSS J115727.61+431806.3}
\begin{figure}
    \includegraphics[width=9cm]{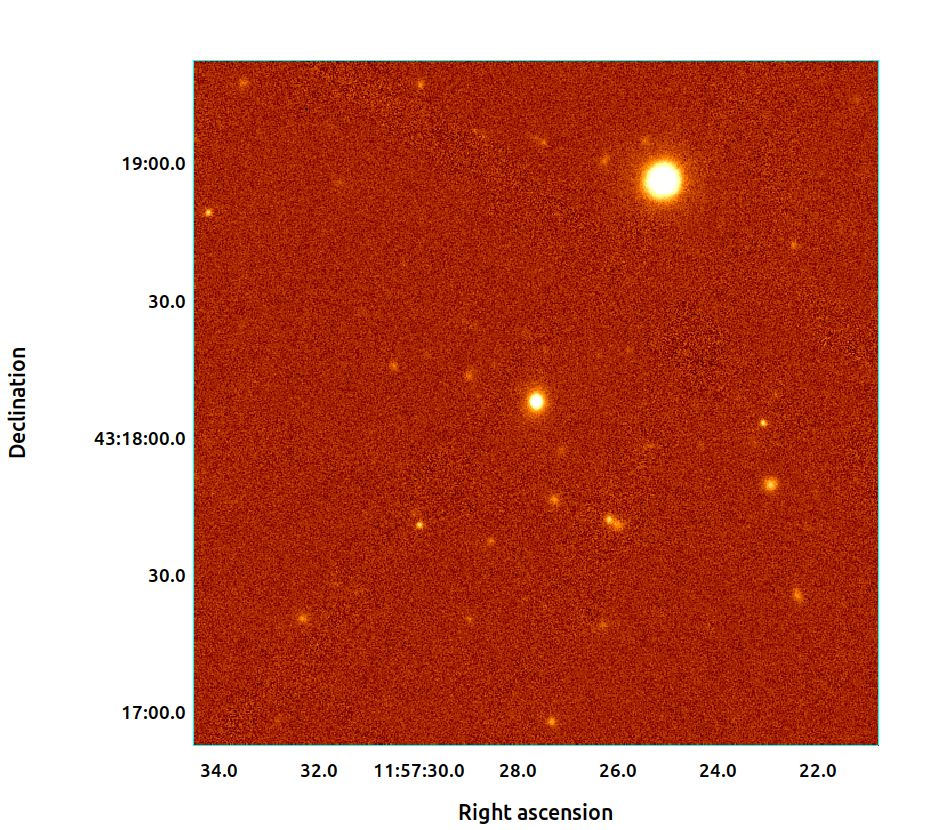}
    \caption{PS1 i-band image of J115727.61+431806.3.}
    \label{fig:image7}
\end{figure}

\begin{figure}
    \centering
    \includegraphics[width=9cm]{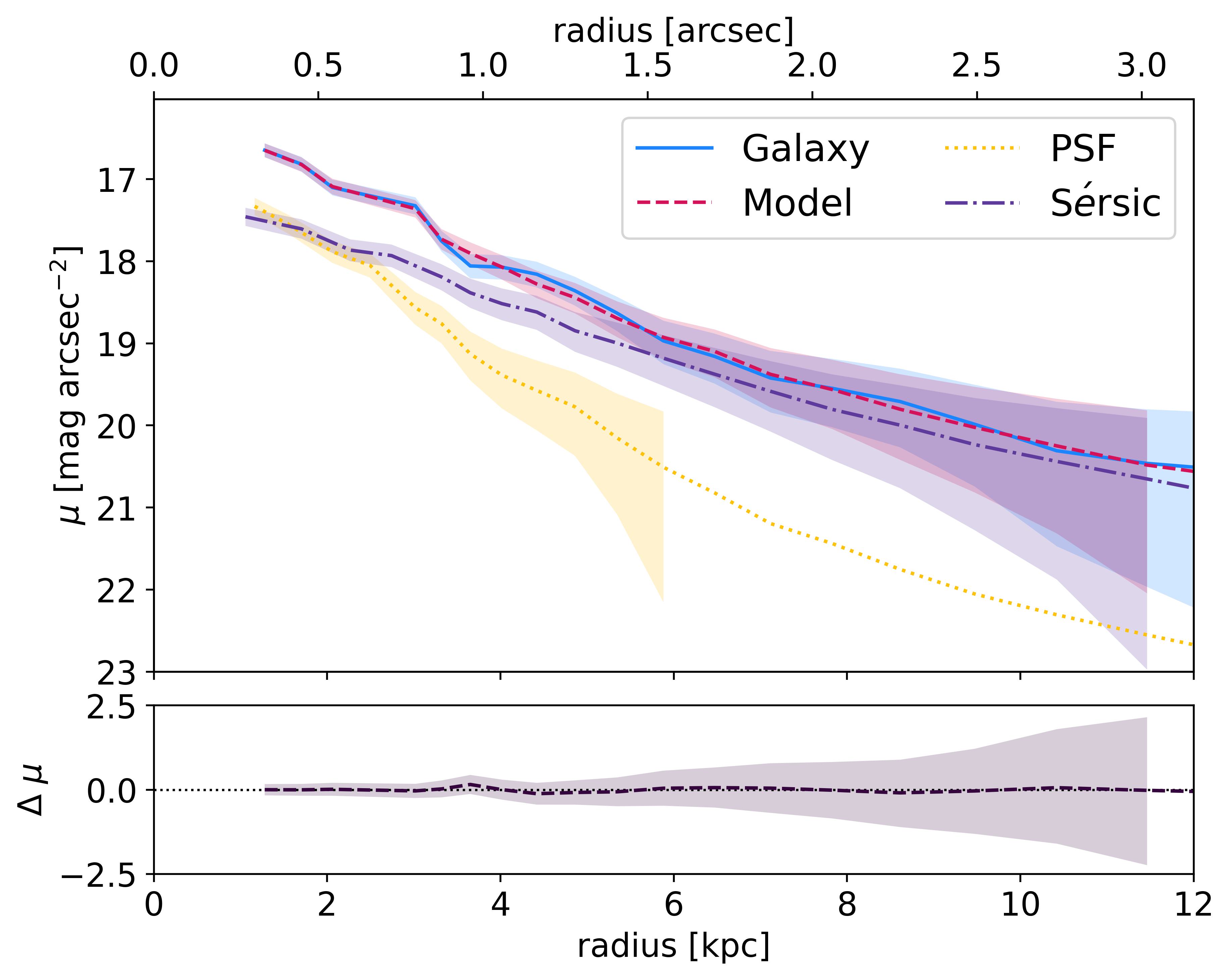}
    \caption{Radial surface brightness profiles of J1157+4318: the observed profile and the model. Legend in the plot. The shaded area around each profile describes the associated errors. The low panel shows the variation in magnitude.}
    \label{fig:host7}
\end{figure}

J115727.61+431806.3 <B3 1154+435>, ($z=$ 0.230, Intermediate AGN) is the only case for which there is no agreement between the single and the multi-band analysis. First of all, it was challenging to retrieve a PSF for the y-band image since the only star, close enough to the target, was very bright. To properly model the PSF, seven Sérsic functions were needed. The host galaxy was then fitted with a PSF and a single Sérsic function for the bulge (parameters can be found in Tab.~\ref{tab:tab1}). The bulge was barely resolved, and its Sérsic index was higher than 4, meaning that this AGN could be hosted in an early-type galaxy. Freezing the coordinates of the Sérsic function instead brought a low Sérsic index, indicating a disc-like host galaxy.
For this source, it was not possible to calculate the positive errors since GALFIT did not converge trying to fit the model plus $\sigma_{sky}$ and so those values were reported as \textit{N/A} in Tab.~\ref{tab:tab1}. Multiband analysis on the $g_{P1}$, $z_{P1}$ (which contains H$\alpha$) and $y_{P1}$ bands gave a different result for the morphology of the host galaxy, finding it as a late-type. We left the total magnitude, for both PSF and the bulge, and the $R_{\rm e}$ of the bulge free to vary, while the Sérsic index of the bulge was allowed to vary as a constant function of the wavelength.
No results were found in the literature on this object and neither we could confirm any kind of morphology also from the i-band image (Fig.~\ref{fig:image7}), it is not possible to retrieve any information about the morphology of this galaxy. Despite the inconsistency between the single- and multi-band analysis, the radial surface brightness profile (Fig.~\ref{fig:host7}) shows that the model of the early-type galaxy well represents the galaxy profile.

\subsection{SDSS J140416.35+411748.7}
\begin{figure}
    \includegraphics[width=9cm]{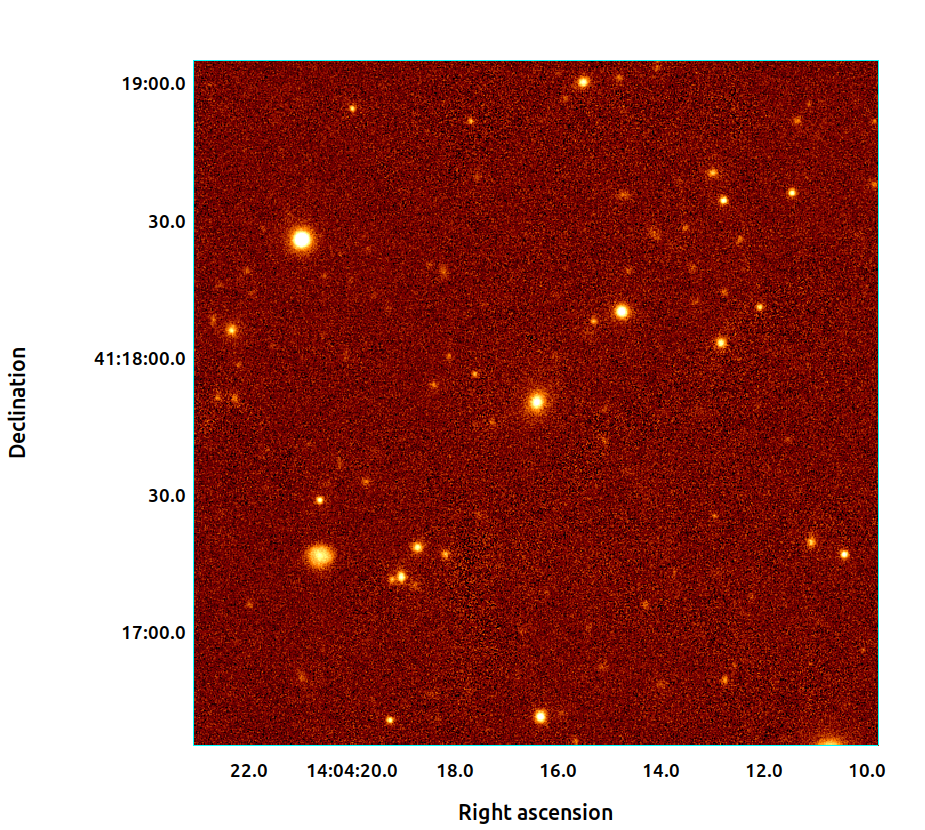}
    \caption{PS1 i-band image of J140416.35+411748.7.}
    \label{fig:image8}
\end{figure}

\begin{figure}
    \centering
    \includegraphics[width=9cm]{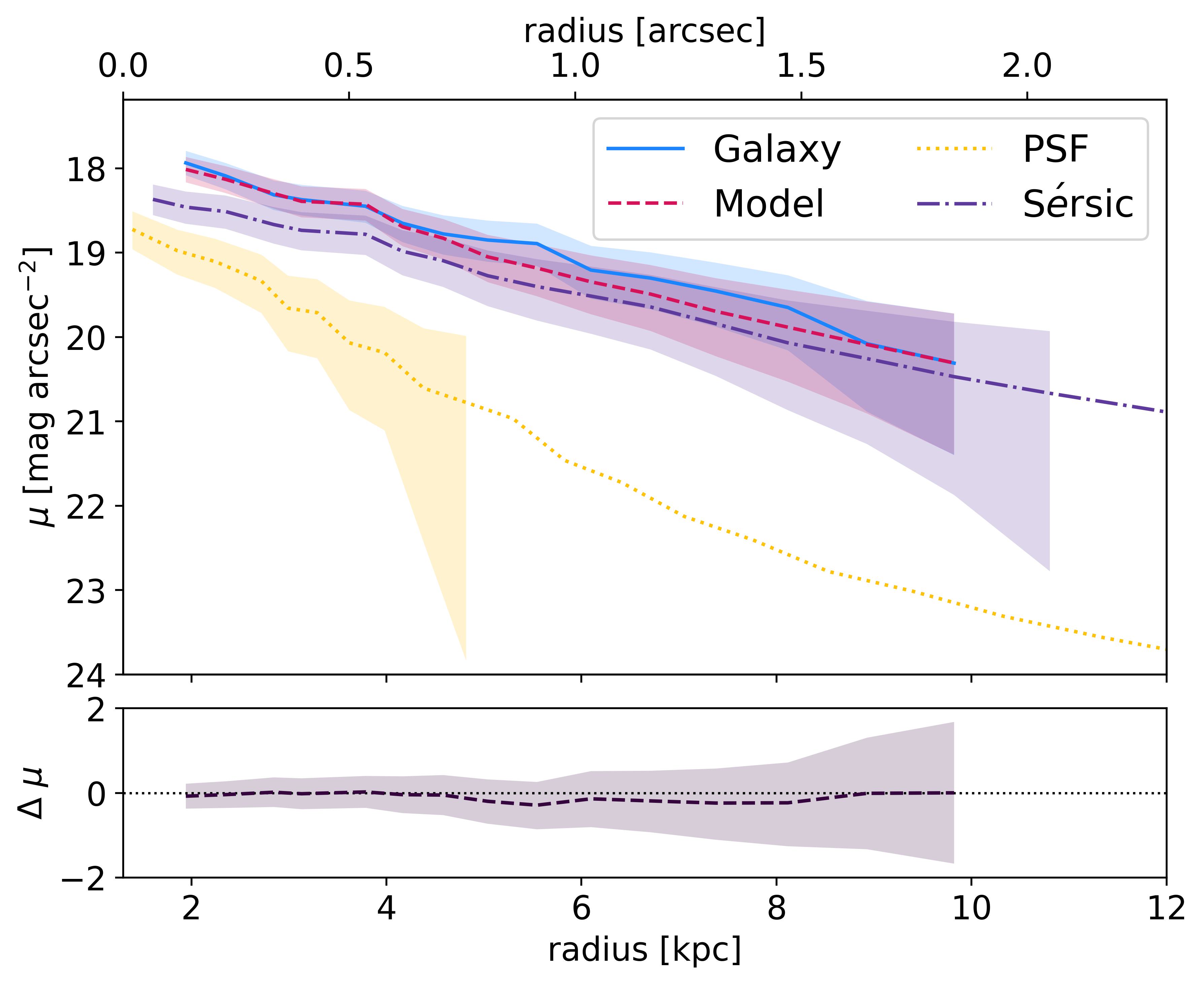}
    \caption{Radial surface brightness profiles of J1404+4117: the observed profile and the model. Legend in the plot. The shaded area around each profile describes the associated errors. The lower panel shows the variation in magnitude.}
    \label{fig:host8}
\end{figure}

Although J140416.35+411748.7 <B3 1402+415>, ($z=$ 0.360) is a Type 2 AGN, it was modelled with a PSF plus a Sérsic function. The PSF model was obtained by fitting 3 Sérsic functions. The parameters of the best fit are shown in Tab.~\ref{tab:tab1}. Again, the low Sérsic index and the $R_{\rm e}$ extending outside the bulge region can be explained by a hybrid component between the bulge and the disc, suggesting anyway a disc-like morphology.
The multi-band analysis made with GALFITM on the $g_{P1}$, $z_{P1}$ (containing H$\alpha$) and $y_{P1}$ band, leaving the total magnitude, the effective radius $R_{\rm e}$ and the Sérsic index free to vary, confirms the classification found with the single-band analysis. No morphological classification has been found in the literature on this object and also in this case it is not possible to retrieve more information from the i-band image Fig.~\ref{fig:image8}, apart from a shallow feature surrounding the galaxy centre and resembling a disc. In Fig.~\ref{fig:host8} it is possible to see how the model reproduces quite well the light profile of the host galaxy of J1404+4117, showing a shallow deviation at large radii.

\subsection{SDSS J140942.44+360415.8} 
\begin{figure}
    \includegraphics[width=9cm]{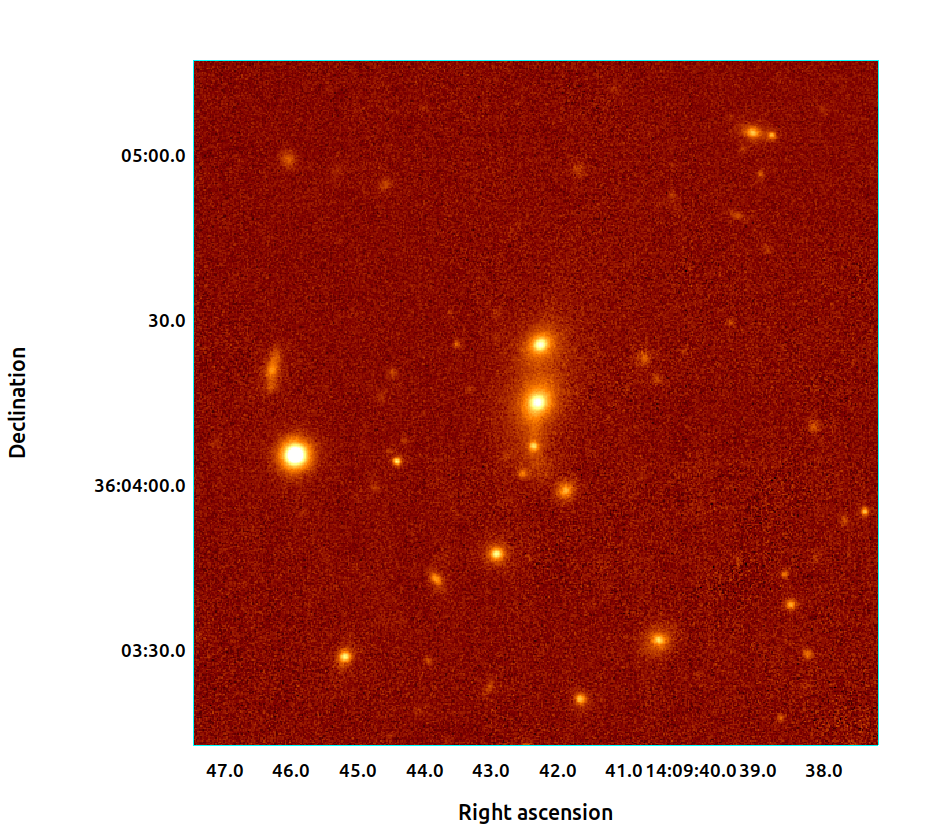}
    \caption{PS1 i-band image of J140942.44+360415.8.}
    \label{fig:image9}
\end{figure}

\begin{figure}
    \centering
    \includegraphics[width=9cm]{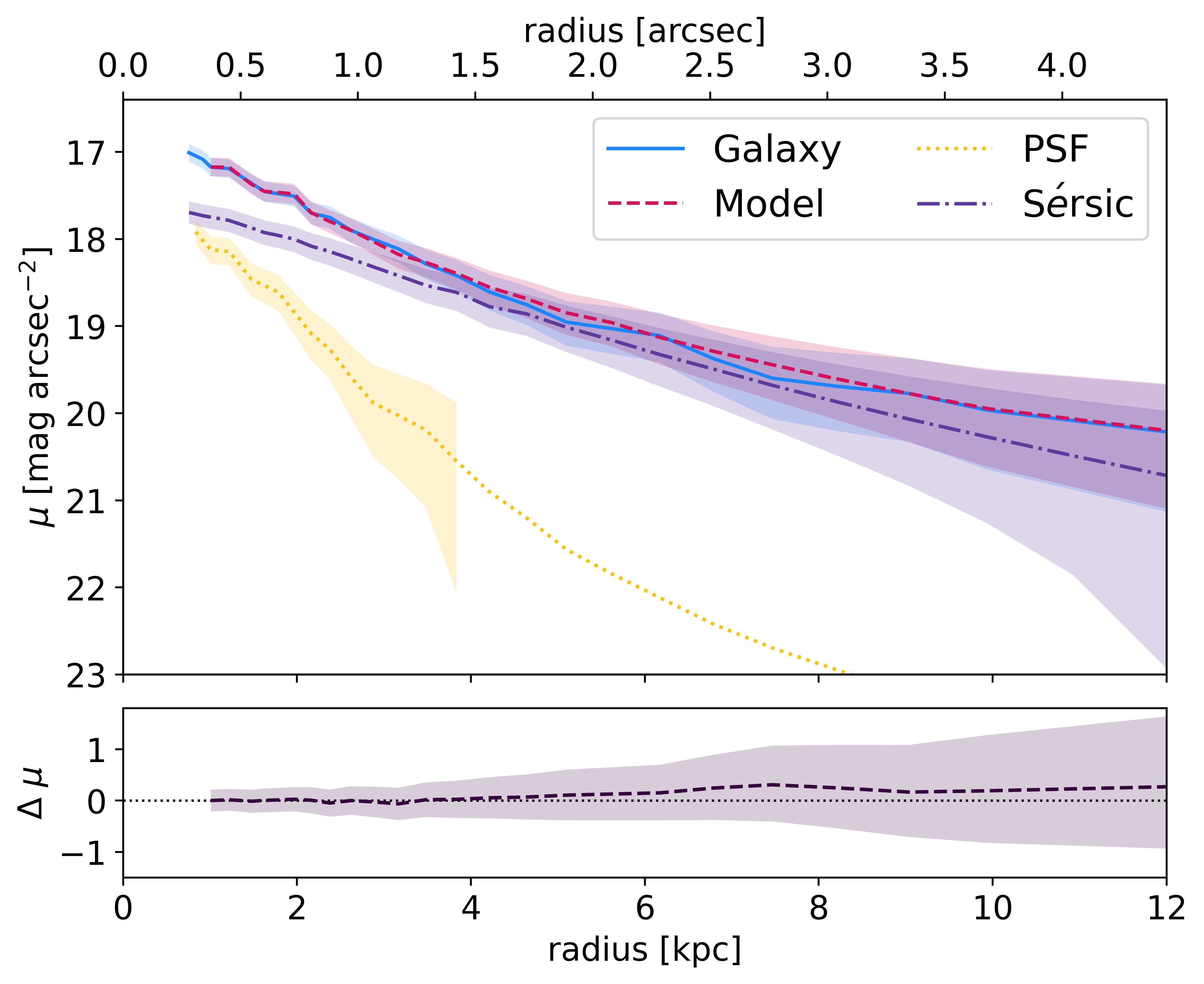}
    \caption{Radial surface brightness profiles of J1409+3604: the observed profile and the model. Legend in the plot. The shaded area around each profile describes the associated errors. The lower panel shows the variation in magnitude.}
    \label{fig:host9}
\end{figure}

The best fit of J140942.44+360415.8 ($z=$ 0.148, Intermediate AGN) was obtained with a PSF (modelled with 3 Sérsic functions) and a Sérsic function representing the hybrid component between the bulge and disc (as explained in the previous subsections; parameters in Tab.~\ref{tab:tab1}). The low value of the Sérsic index leads to the classification of a late-type galaxy for this host galaxy. The GALFITM analysis, for the $g_{P1}$, $i_{P1}$ (which contains H$\alpha$) and $y_{P1}$ bands, with all parameters free to vary, converges to a Sérsic index of almost 2 for all three bands. The i-band image of the source, shown in Fig.~\ref{fig:image9}, confirms the late-type features for this galaxy, showing a disc morphology. Despite this, the source has been visually classified as an elliptical by \citet{Nascimento22}. 
Fig.~\ref{fig:host9} shows the radial surface brightness profile of the source, and the galaxy profile is well represented by the model made up of the PSF plus the Sérsic function.

\subsection{SDSS J164311.34+315618.4}
\begin{figure}
    \includegraphics[width=9cm]{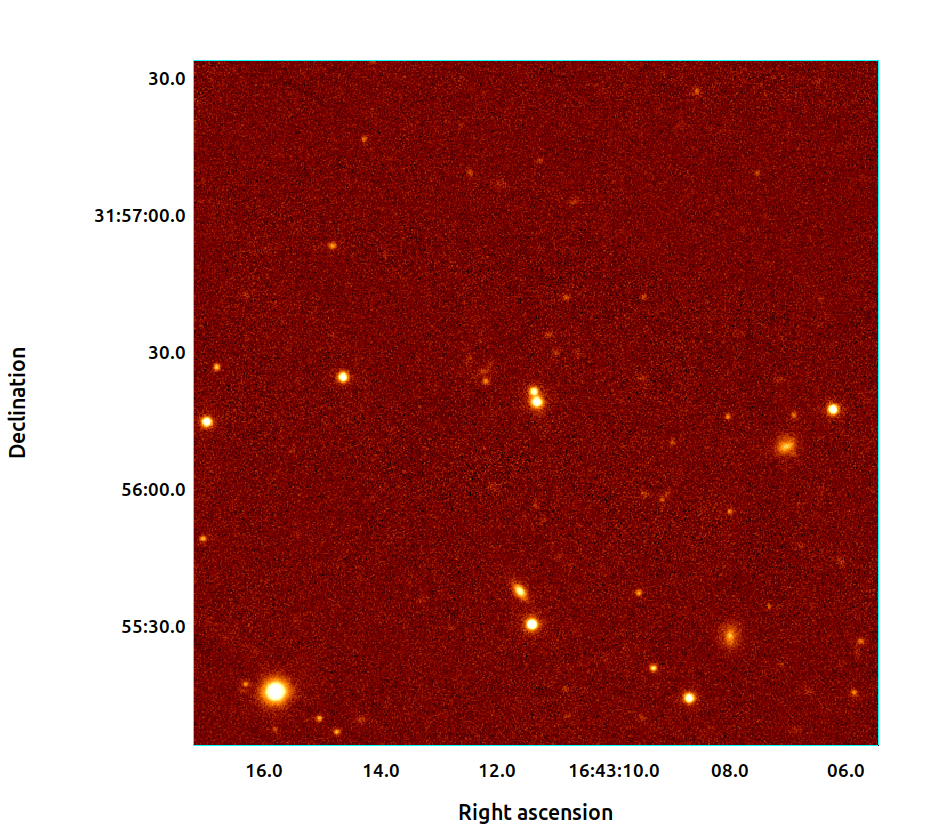}
    \caption{PS1 i-band image of J164311.34+315618.4.}
    \label{fig:image10}
\end{figure}

\begin{figure}
    \centering
    \includegraphics[width=9cm]{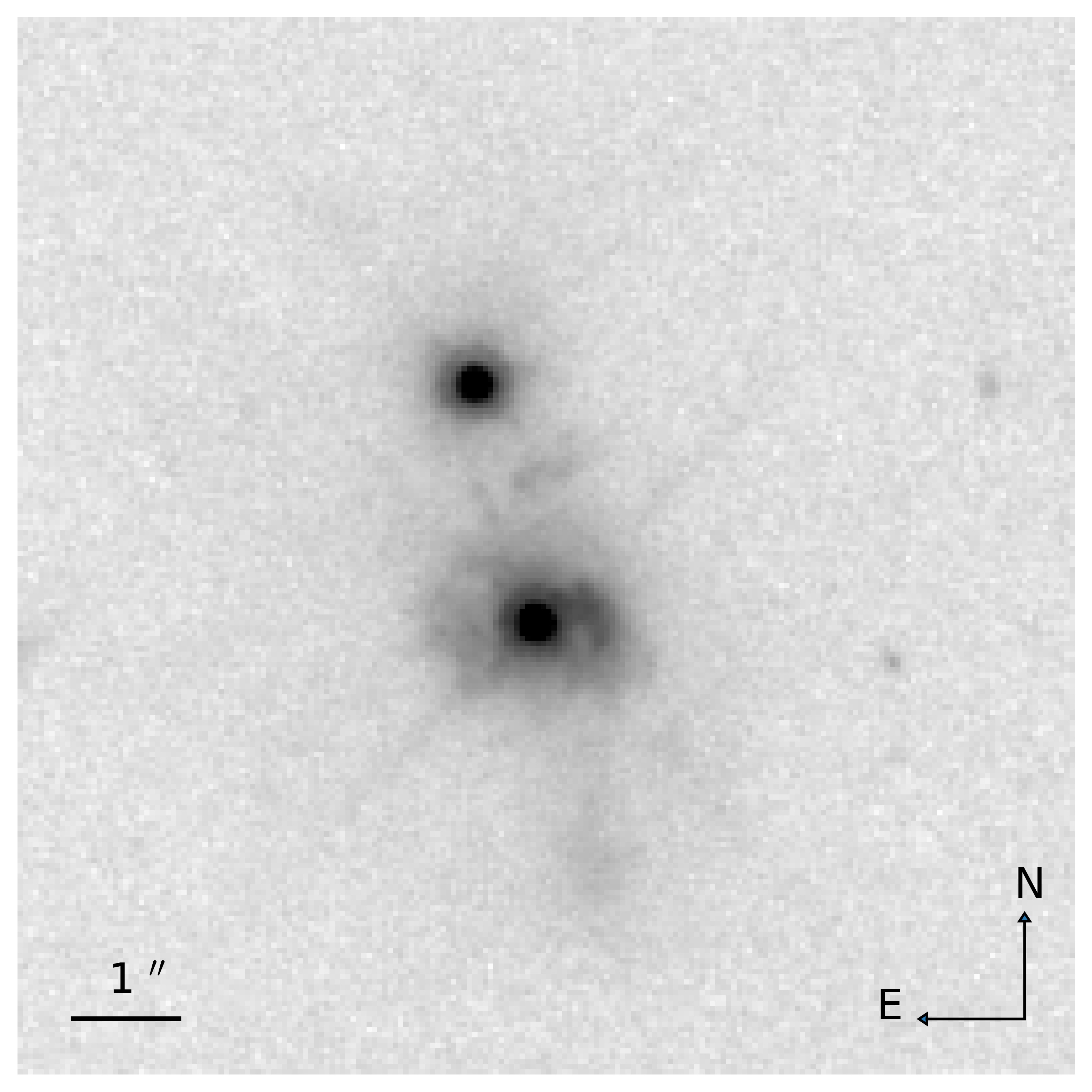}
    \caption{Image of J1643+3156 and its companion, taken with the Wide Field Channel of the ACS of \textrm{HST} in the F814W filter. It is clear that the host galaxy of this LLC (the object at the centre of the image) does not resemble an elliptical galaxy, showing an halo and features of a possible past interaction.}
    \label{fig:host10hubble}
\end{figure}

\begin{figure}
    \centering
    \includegraphics[width=9cm]{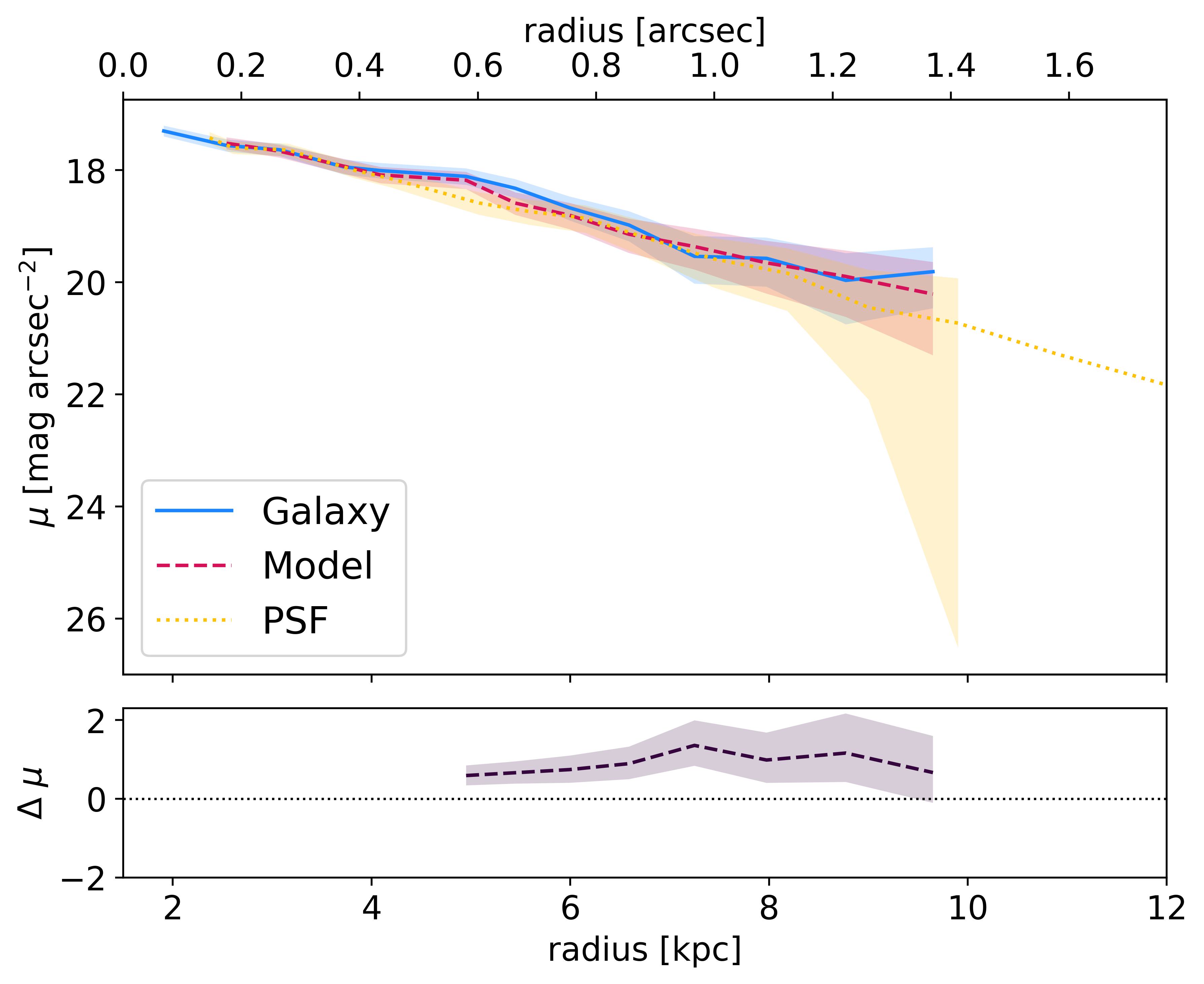}
    \caption{Radial surface brightness profiles of J1643+3156: the observed profile and the model. Legend in the plot. The shaded area around each profile describes the associated errors. The lower panel shows the variation in magnitude.}
    \label{fig:host10}
\end{figure}

J164311.34+315618.4 ($z=$ 0.586, Intermediate AGN) is part of an interacting system, a merging of two AGN, as shown in Fig.~\ref{fig:image10}, the i-band image of the source. The binary quasar associated with this radio source has been classified by \citep{Brotherton99} based on optical observations. Its disturbed radio morphology has been discussed by \citet{Kunert11}, where the
\textrm{HST} images (\citet{Martel05}) are shown. From the \textrm{HST} image (Fig.~\ref{fig:host10hubble}, a zoom-in of the entire image on the binary quasar) it appears clear that this source is not hosted in an elliptical galaxy, but rather in an irregular or spiral galaxy. 
The best fit for the entire system was obtained instead with two PSFs, one for each object, and the best PSF model of the y-band image was obtained with a single Sérsic function. The goodness of fit was based on the reduced $\chi^{2}_{\nu}$= 1.151$_{-1.023}^{+1.042}$. The PSF magnitude of the fit was PSF$_1$: mag = 16.59$_{-0.12}^{+0.13}$. PSF$_2$: mag = 17.42$_{-0.25}^{+0.34}$, respectively. 
In this case, nothing can be said about the morphology of this galaxy, and nothing has been found in the literature. We used GALFITM on the $g_{P1}$ and $y_{P1}$ bands (containing H$\alpha$), leaving all parameters free to vary, confirming that a single PSF was enough to fit the light profile of this source.  The radial surface brightness profile model (Fig.~\ref{fig:host10}) deviates a bit from the galaxy profile, probably due to the presence of the second source, indeed this effect is more prominent at large radii.

\section{Discussion}
\label{sec:discussion}
\subsection{The host galaxy of LLCs}
\renewcommand{\arraystretch}{1.5}
\begin{table*}[ht!]
\caption[]{Morphological classification from this analysis and comparison with results found in literature.}
\centering
\begin{tabular}{l l l }
\hline\hline
Source   &  Morphology & Literature \\ 
\hline\hline
J0028+0055  & Late-type with pseudo-bulge & Spiral$^a$ \\
J0757+3959  & Late-type with pseudo-bulge & Spiral$^a$ \\
J0848+0136  & Late-type & Early-type$^b$\\
J0926+0745  & No classification & No result \\
J0945+3521  & Late-type with possible pseudo-bulge & Irregular/Merger$^a$\\
J1143+0535  & No classification & Irregular/Merger$^a$\\
J1157+4318  & Uncertain classification & No result\\
J1404+4117  & Late-type with possible pseudo-bulge & No result\\
J1409+3604  & Late-type with possible pseudo-bulge & Elliptical$^a$\\ 
J1643+3156  & No classification & No result\\ \hline

\end{tabular}
\tablefoot{Columns: (1) Short name; (2) Result from this analysis; (3) Morphological classification found in the literature: references (a) \cite{Nascimento22}, (b) \cite{Urbano18}.}
\label{tab:morph}
\end{table*}

In this work, we have performed an analysis of the host galaxy morphology of the LLCs sample used to study the parent population of jetted F-NLS1s \citep{Berton16c}. Our morphological classification is based on the 2D analysis carried out with GALFIT and also with GALFITM, and in turn, when possible, on the light profile of the galaxy bulge. The bulge is of paramount importance to understand the formation and evolutionary processes that occurred throughout the history of the galaxy. Generally, large bulge Sérsic indexes are associated with massive galaxies; indeed, the galaxy' size and luminosity correlate with its value. In elliptical galaxies, light typically follows a de Vaucouleurs profile with Sérsic index $n_{\rm b} = 4$. Classical bulges are usually associated with past violent events such as merging \citep{Gadotti09a, Gadotti09b}. This can lead to the disruption of discs in general (\cite{Toomre72}, \cite{Barnes96}), and of pseudo-bulges, which are instead formed via secular evolution processes or minor mergers \citep{Orbandexivry11, Jackson22}. Only in the last 20 years has the importance of secular processes, minor mergers and gas accretion from disc and bar instability been recognised \citep{Kormendy04, Steinborn18} as fuel to the AGN activity (e.g., \cite{ho09, Cisternas11, Villforth17}) and their role in the co-evolution between host galaxies and their supermassive black holes (\cite{Parry09,Martin18,McAlpine20}).
For our purposes, we classify as late-type those galaxies hosting a pseudobulge, i.e. with n$_b \approx$ 2. Conversely, early-type galaxies have $n_{\rm b} \geq 4$. In between, the morphology can be somewhat mixed, and it is not wise to draw clear conclusions about the host without additional deeper observations.\\
It is known that precise morphological classification of galaxies requires well-resolved images, regardless of redshift (\cite{Davari17}, \cite{Zhuang22}). At these distances (the farthest source is almost at $z$ $\approx$ 0.6), the physical scale can reach the value of 6~kpc arcsec$^{-1}$ in radius, and considering the resolution of the PS1 images (0.258 arcsec/pix), a reliable bulge-to-disc decomposition can be hard to obtain.
Following these considerations, we decided to proceed with a single Sérsic component model (in addition to the PSF when needed) to decompose the galaxy light profile under the PS1 survey conditions. Only in two cases, indeed for the closest sources, another Sérsic function was needed to obtain the best fit. 
The results from our morphological analysis and those found in the literature are listed in 
Tab.~\ref{tab:morph}.\\
We found that six out of ten sources from our sample are definitely hosted in late-type galaxies. The presence of pseudo-bulges is verified for the two sources (J0028+0055, J0757+3959) which have been decomposed with two components. For the other late-type galaxies, as previously explained, the boundary of the single component model brought to an hybrid component with a high $R_{\rm e}$ and a low Sérsic index, subtending the presence of pseudo-bulges.
For three of the ten sources (J0926+0745, J1143+0535, J1643+3156) it was not possible to retrieve any morphological classification since their light profiles were fit only with a PSF; for one (J1643+3156) of these three the \textrm{HST} image clearly shows the presence of a disc in the host galaxy. 
In the end, only in one case (J1157+4318 ) there was no agreement between the single- and the multi-band analysis, so we cannot confirm a unique morphological type for its host galaxy. The early-type classification, provided by the single-band analysis, is consistent with the radial surface brightness profile shown in the plot.\\
When it was possible to compare our results with the literature, we found the same morphological classification in the half of the cases Tab.~\ref{tab:morph}. This comparison cannot be used as a confirmation or denial of the validity of our results, since it is only applied to a few sources. Also, the methodology used by \citet{Nascimento22} for morphological analysis is visual inspection, which could lead to misleading results.
Overall, our results confirm the main idea that CSS sources in general, and LLCs in particular, can be found to be hosted not only in large elliptical galaxies but also in late-type galaxies, even with pseudo-bulges. 

\subsection{Multi-band approach}
As far as we know, only three studies \citep{Zhuang22, Martorano23, Acharya24} focused on the host galaxy morphology of AGN, obtained with a multi-band approach using GALFITM, can be found in the literature. Despite the well-known wavelength dependence of the galaxy structure, until a few years ago this aspect had not been taken into account in the AGN host galaxy decomposition studies. As previously discussed, the parameters defining the galaxy can vary with wavelength since they are affected by the different stellar populations, the presence of the dust and the metallicity gradient, which in turn are wavelength dependent. The same kind of dependence can be shared by the Sérsic index. As a consequence, also the morphology of a galaxy can depend on the band in which it is defined. For example, late-type galaxies tend to be brighter at shorter wavelengths because of the disc emission and at longer wavelengths because of the prominence of the bulge. \citet{Kelvin12}
found that in both early- and late-type galaxies, the half-light radius and the Sérsic index show a smooth and systematic variation with the wavelength: from the ultraviolet (UV) to the near-infrared (NIR), the former decreases while the latter increases. This can be explained by the longer wavelengths being more able to trace older stars, while the shorter ones are more sensitive to younger stellar populations. 
We decided to search for this kind of relations and, at the same time, to use the multi-band approach as proof of the validity of the results, especially since the limitation due to the single component model. 
We found a unique classification for all of our sources, except for one. The mild difference in the Sérsic index and in $R_{\rm e}$, a smooth increase and decrease moving from the $g_{P1}$ to the $y_{P1}$ filter, respectively, confirm the trend shown in the work of \citet{Kelvin12}. This kind of variation was never so extreme to completely change the morphological type defined with the single-band analysis. This also confirms the results of \citet{Martorano23}, in which the authors claimed a weak change in the Sérsic index with wavelength. It is also worth noticing that the wavelength range covered by the PS1 filters goes from optical to NIR, without reaching the UV. This can also affect the kind of variation we can see in the Sérsic index.
Furthermore, in the case of J1157+4318, we do not link the different morphological classification obtained from the single- and the multi-band analysis with the wavelength dependence of the Sérsic index. This inconsistency can be instead attributed to the difficulty in creating the galaxy model itself from the single-band image.

\subsection{The link with NLS1s}
By definition, GPS and CSS have a threshold in their radio power (log$P_{1.4} > 25$ W Hz$^{-1}$, \citealp{Odea98}), therefore they are rather bright objects. It is known that the power of relativistic jets correlates nonlinearly with the black hole mass \citep{Heinz03, Foschini14}. Due to the presence of well-known scaling relations between the galaxy and the mass of its central black hole \citep{Ferrarese00}, any study focused on the host galaxy of these objects will be skewed toward large ellipticals. On the other hand, LLCs do not have such power thresholds, and they can be powered by low-mass black holes, as proved by \citet{Berton16c}, similar to those harboured by jetted NLS1s. 

The main finding of this paper, i.e. that the host galaxies of LLCs can have a late-type morphology, has robust implications on the nature of the parent population of $\gamma$-ray emitting (or simply relativistically beamed) NLS1s.\\
There are also other kind of low-luminosity sources, like FR type 0 (see \citet{Baldi23} for a recent review), that have been observed in the $\gamma$-rays (e.g. \citet{Pannikkote23}) and that at the same time show a flat radio spectrum instead, likely being inclined at small angles. This underlines the importance of the precise measurements of the radio spectral index.\\
In the model proposed by \citet{Berton16c} and \citet{Berton17}, the radio luminosity function of LLCs was compared with that of beamed F-NLS1s, finding that LLCs may be jetted F-NLS1s observed at high inclination, often with Type 2 optical spectra. One of the potential issues with that model \citep{Berton16c} was in fact the potentially different host galaxy morphology. For jetted NLS1s, the overwhelming majority of hosts are late-type, as proven by a number of works in the literature \citep[e.g.,][]{Olguiniglesias20}. Our results indicate that LLCs share similar characteristics with NLS1s, since at least six of their hosts are late-type as well, with only one source for which the morphological classification was unclear. Therefore, the orientation-based unification model seems to be supported by our findings.  Also, the radio luminosity range (1.4 GHz) of this sample of LLCs, lies in the wider range of radio luminosity for F-NLS1s (and steep spectrum NLS1s), found by \citet{Singh18} for the biggest sample of NLS1s with enhanced radio emission, confirming the parent population scenario.

The general conclusion we can draw about the parent population of beamed NLS1s is that several diverse objects can be part of it, including LLCs. In general, when observed at large angles, beamed NLS1s can alternatively look like NLS1s with an extended radio jet, although this seems to be a rare occurrence, or as relatively weak and young radio sources, such as LLCs, with a Type 2/Intermediate optical spectrum. At the same time, several misaligned jetted NLS1s are definitely hidden among objects with no known radio emission. This is proved by the discovery of extremely variable radio emission \citep{Jarvela23} in this class of AGN \citep{Lahteenmaki18, Berton20b, Jarvela21}, but it is also worth noting that the lack of relativistic beaming would make even $\gamma$-ray NLS1s barely detectable at radio frequencies \citep{Berton18a, Jarvela22}. Additional studies, especially using optical spectropolarimetry \citep{Antonucci85}, are needed to fully understand this issue and identify the high-inclination counterparts of jetted NLS1s. \\

\subsection{The literature issue}
As well explained in the recent review on CSS and GPS \citep{Odea21}, it is known that generally PS and CSS sources are found to be hosted in bright elliptical galaxies populated by old stellar populations. Indeed, very recent works (e.g. \cite{Gordon23}) strongly support this statement. However, \citet{Odea21} also underline that
several sources have also been found in late-type galaxies with strong disc components, which still seems to be an exception from the general rule.\\
Our concern, which also explains the reason behind this kind of analysis, is to step back from this statement, validating the scenario by which early-type galaxies are not the only ones capable of hosting these radio sources, and maybe neither the majority of them.\\
In the past, several studies have been dedicated to the host galaxy of PS. One of the larger samples is that of \citet{Devries00}, who found that the hosts of GPS, CSS, and FR II radio galaxies are consistent with each other. Their results show very often Sérsic indexes well below 4, and in some cases $\leq$ 2, suggesting that CSS can be found to be hosted in giant ellipticals as well as in spiral galaxies. This can appear contradictory to what the authors claimed in the end, saying that the host galaxies of their radio sources are found to be regular giant elliptical galaxies, as proved by the absolute magnitudes and surface brightness profiles. 
Later studies started to notice that previous works, suggesting the host galaxies of CSS/GPS sources as passive elliptical galaxies, included objects hosting young stellar populations or showing evidence for recent mergers or interactions \citep{Holt09, Stockton07, Emonts16}.
\citet{Sadler16}, for example, who focused their analysis on Wide-field Infrared Survey Explorer (WISE) survey data of GPS and CSS sources, found a heterogeneous populations in terms of host galaxy type (67$\%$ early-type galaxies, 33$\%$ late-type systems). \citet{Kuzmicz17}, instead, showed that from the WISE colour-colour diagram the 67$\%$ of radio galaxies with recurrent jet activity reside in the region typical for late-type galaxies with ongoing star formation or spiral galaxies. \citet{Nascimento22} instead, from a sample of 58 CSS/GPS, classified 15 as ellipticals, 18 as spirals, 12 as irregular/mergers and 13 as point sources, underlining that PS sources do not show a preferred morphological type of host galaxies.
We also need to keep in mind, as underlined by \citet{Odea21}, that galaxies with a lower bulge/disc ratios \citep{Pierce19} and/or with less luminous stellar bulges \citep{Vaddi16} tend to host AGN with lower radio power, such as LLC sources. Those features are indeed usually found in late-type galaxies.
In the end, last year \citet{Duggal24} found three out of seven CSSs of their sample to be hosted in strong spiral or disc-like galaxies, again claiming this as an interesting result, since very few known sources have been found in late type hosts (\cite{Heckman82, Johnston10, Anderson13}.
Our analysis, the first one focused on the LLCs hosts, can be located in this framework validating the late-type galaxies as hosts of PS sources.
Coming back to what we were discussing at the beginning of this paragraph, it is clear how despite the multiple evidence of spiral galaxies hosting PS sources, nowadays this is still considered as an interesting result.
Reporting all those studies, we do not aim to point them as being incorrect for their claims, rather we aim to underline how certain kind of statement can strongly affect following studies, even when several proofs against the validity of those statements have been published. General rules can be changed by experimental evidences.\\
A common theme of all these works is the necessity of expanding the sample, which can bring to a reliable statistic on the host galaxy population of radio sources, helping in defying a possible new scenario in the literature. 

\section{Summary and conclusions}
\label{sec:summary}
This is a pilot study with a relatively small sample and with constraints given by the use of PS1 images, and it confirms that LLCs are hosted in disc-like galaxies, as NLS1s. Starting from the LLCs sample of \citet{Berton16c}, we performed the photometric decomposition of their PS1 images in all \textit{grizy}$_{P1}$ filters, using the 2D fitting algorithm for single- and multi-band analysis, GALFIT and GALFITM. We found six out of ten sources to be hosted in late-type galaxies, probably with pseudo-bulges, three point sources, and one of undefined classification. Except for one case, the morphological classification does not vary with wavelength, and the Sérsic indices only show mild changes within the different bands, as expected. This finding adds an important piece to the puzzle of NLS1s, recognising LLCs as their parent population, confirming the hypothesis made in \citet{Berton16c}. 
To our knowledge, this is the only statistically complete sample (following the criterion in \cite{Berton16c}) of LLCs available in the literature and the only one for which a morphological analysis of the host galaxies has been conducted. Several works have been focused on the classification of the morphological type of CSS host galaxies in general, but none on LLCs in particular. 
As a first step, a new and larger complete sample of LLCs has to be defined. Then, facilities providing high-resolution data will be needed to conduct a comprehensive analysis of the black hole masses, the Eddington ratio distributions and the morphology of the host galaxies of LLCs and F-NLS1s, to definitely confirm the parent population. This can be achieved using ground-based facilities, such as the ones provided by the European Southern Observatory (ESO) and the Very Large Telescope in particular. These telescopes, working in the optical range, can simultaneously offer high signal-to-noise spectra and images with enough resolution for the bulge-to-disc decomposition. 
Additionally, space facilities, as the state-of-the-art instruments aboard the \textit{James Webb Space Telescope} and the \textrm{HST}, are well-suited for this purpose, providing unique spatially resolved structural information on galaxies in the near-infrared and optical bands, respectively.

\begin{acknowledgements}
A.V. and M.B. acknowledge the support from the ESO Early-Career Scientific Visitor Programme. 
M.K.B. acknowledge support from the “National Science Centre, Poland”
under grant No. 2017/26/E/ST9/00216. The Pan-STARRS1 Surveys (PS1) and the PS1 public science archive have been made possible through contributions by the Institute for Astronomy, the University of Hawaii, the Pan-STARRS Project Office, the Max-Planck Society and its participating institutes, the Max Planck Institute for Astronomy, Heidelberg and the Max Planck Institute for Extraterrestrial Physics, Garching, The Johns Hopkins University, Durham University, the University of Edinburgh, the Queen's University Belfast, the Harvard-Smithsonian Center for Astrophysics, the Las Cumbres Observatory Global Telescope Network Incorporated, the National Central University of Taiwan, the Space Telescope Science Institute, the National Aeronautics and Space Administration under Grant No. NNX08AR22G issued through the Planetary Science Division of the NASA Science Mission Directorate, the National Science Foundation Grant No. AST-1238877, the University of Maryland, Eotvos Lorand University (ELTE), the Los Alamos National Laboratory, and the Gordon and Betty Moore Foundation.\\
We thank Boris H\"au\ss ler for valuable advice and explanation on the use of GALFITM. We thank Lorenzo Cavallo and Edoardo Borsato for the encouragement in using GALFITM and for all the advices and tips aimed at improving the plots/image graphical design. This research has made use of the NASA/IPAC Extragalactic Database (NED), which is operated by the Jet Propulsion Laboratory, California Institute of Technology, under contract with the National Aeronautics and Space Administration.
\end{acknowledgements}


%
\bibliographystyle{aa} 
\bibliography{./main.bib} 
%

\end{document}